\documentclass{aastex61}

\newcommand\aastex{AAS\TeX}

\usepackage[figuresright]{rotating}
\usepackage{graphicx}
\usepackage{array}

\shorttitle{\aastex\ Binding Energies of Interstellar Species}
\shortauthors{Das et al.}

\begin{document}

\title{An Approach to Estimate the Binding Energy of Interstellar Species}
\correspondingauthor{Ankan Das}
\email{ankan.das@gmail.com}

\author{Ankan Das}
\affiliation{Indian Centre for Space Physics, 43 Chalantika, Garia Station Road, Kolkata 700084, India}
\author{Milan Sil}
\affiliation{Indian Centre for Space Physics, 43 Chalantika, Garia Station Road, Kolkata 700084, India}
\author{Prasanta Gorai}
\affiliation{Indian Centre for Space Physics, 43 Chalantika, Garia Station Road, Kolkata 700084, India}
\author{Sandip K. Chakrabarti}
\affiliation{S. N. Bose National Centre for Basic Sciences, Salt Lake, Kolkata 700106, India}
\affiliation{Indian Centre for Space Physics, 43 Chalantika, Garia Station Road, Kolkata 700084, India}
\author{J. C. Loison}
\affiliation{Institute of Molecular Sciences, University of Bordeaux, CNRS UMR 5255, France }

\begin{abstract}
One of the major obstacles to accurately model the interstellar chemistry is an inadequate knowledge
about the binding energy (BE) of interstellar species with dust grains.                 
In denser region of molecular cloud, where very complex chemistry is active,   
interstellar dust is predominantly covered by H$_2$O molecules and thus it is essential to
know the interaction of gas phase species with water ice to trace realistic physical and chemical processes.
To this effect, we consider water (cluster) ice to calculate the BE of several atoms, molecules,
and radicals of astrochemical interest. Systematic studies have been carried out to
come up with a relatively more accurate BE of astrophysically relevant species on water
ice. We increase the size of the water cluster methodically to capture the realistic situation.
Sequentially one,  three, four, five and six water molecules are considered to represent water
ice analogue in increasing order of complexity. We notice that for most of the species considered here, 
as we increase the cluster size, our calculated BE value starts to converge 
towards the experimentally obtained value. More specifically, our computed results
with  water c-pentamer (average deviation from experiment $\sim \pm 15.8 \%$) and c-hexamer (chair) 
(average deviation from experiment $\sim \pm 16.7 \%$) configuration are found to be more nearer  
as the experimentally obtained value than other water clusters considered.
\end{abstract}


\section{Introduction}
Molecules in space are synthesized via gas-phase reactions as well as the reactions occurring on 
interstellar grain surfaces. In the process of chemical enrichment, gas and grains 
continuously exchange their chemical components with each other. 
Interstellar dust act similar to a catalyst \citep{garr06,herb09}
during the chemical enrichment of the interstellar medium (ISM). Knowledge of the binding energies (BE) of 
the interstellar species is very crucial to understand synthesis of molecules in the gas phase as well as 
interstellar icy grain mantle. Species which are being produced or trapped on 
interstellar ice may transfer to gas phase by various
desorption mechanism such as non-thermal reactive desorption \citep{garr07} (efficient desorption 
mechanism at low temperature), thermal desorption (efficient at high temperature) and 
energetic processes such as, direct or indirect photo evaporation by photon or cosmic ray particles.
Interestingly, all the parameters causing  desorption are directly or indirectly related to the 
BE of the species \citep{gora17a,gora17b,sil18}. According to \cite{mini14}, the efficiency of non-thermal 
chemical desorption mechanism basically depends on four parameters: enthalpy of formation, degrees of freedom, 
BE and mass of newly formed molecules. \cite{mini16} proposed a new chemical desorption 
equation by relating the equipartition of energy. Recently, \cite{wake17} studied the efficiency 
of chemical desorption for a new set of BEs by including both the chemical 
desorption rates proposed by \cite{garr07} and \cite{mini16}. \\

Since 1990s Temperature Programmed Desorption (TPD) technique is used to experimentally determine the BE values. Although TPD measures the
desorption energy, this energy essentially is the binding energy of the species if there are no activated processes. Experimentally 
determined BE depends on the nature of substrate from which the species desorb and on several other parameters like the property of
the deposited ices (i.e., pure, mixed, or layered). These parameters has an effect on the obtained BE values from TPD experiments.
But with the TPD method, it is quite difficult to provide the BE values of the radicals because of their short life under the 
laboratory conditions. However, above $40$ K, the mobility of the radicals exponentially increased
and can actively take part in controlling the surface chemistry. In order to map the chemical composition in the intermediate temperature
($40-80$ K) regime, it is essential to have an estimation of the BE of these radicals.
To this effect, parametric computational studies can provide faster 
information as compared to experimental studies. Prior to the era of TPD experiments, some 
estimated values of BE were used in gas-grain chemical models \citep{tiel82,hase93,char97}. 
These estimations were based on the polarizability of molecules or atoms, which provides an estimate of  
strength of van der Waals interaction with a bare grain surface.
Silicate and carbonaceous type grains are abundant in the ISM. However, denser regions
of molecular clouds are predominantly composed of H$_2$O in amorphous phase with addition
of some other impurities, such as CO, CO$_2$, NH$_3$, CH$_4$, H$_2$CO and CH$_3$OH etc. 
Though early experiments \citep{chak98} and astronomical observations \citep{malf98,mald03} 
clearly suggest the presence of crystalline ice, vapor-deposited 
amorphous ice (also called amorphous solid water, ASW) continues to attract more 
fundamental researchers due to its occurrences in astronomical environments such as icy satellites, comets,
planetary rings and interstellar grains etc. Various surface processes such as adsorption, diffusion, 
tunneling reactions, and nuclear-spin conversion on interstellar ASW are summarized in
\cite{hama13}. ASW is by far the largest component of the icy mantles, with 
abundances of $\sim 10^{-4}$ with respect to the total hydrogen
\citep{will02}, equivalent to coverages of up to $100$ mono-layers (ML). The extinction
threshold ($A_V$) for H$_2$O mantle is $\sim 3.3$ mag \citep{whit88}. 
It should be kept in mind that interstellar ices are thought to have low levels of 
porosity, as they are continuously exposed to external radiation \citep{palu06}. \\

The BE or adsorption energy of various species are the essential input parameters for interstellar 
gas-grain chemistry but only a few have been obtained experimentally.
Among the experimentally obtained BE values, most of the values were
obtained from TPD experiments. 
\cite{coll04} presented an extensive TPD study for a collection of $16$ astrophysically relevant molecular 
species. \cite{ward12} used TPD experiments coupled with time-of-flight mass spectrometry to determine the yield
of OCS and additionally yields value for the computation of desorption energies of O atom and OCS.
The interaction and auto-ionization of HCl on low-temperature
($80-140$ K) water ice surfaces has been studied by \cite{olan11} by using low-energy ($5-250$ eV)
electron-stimulated desorption (ESD) and temperature programmed desorption (TPD).
\cite{duli13} also performed TPD experiment to derive the BE of different species on the silicate substrate.
\cite{nobl12} present experimental study of the desorption of CO, O$_2$ and CO$_2$ from three different surfaces:
non-porous ASW, crystalline ice and amorphous olivine-type silicate. 
Very recently, \cite{pent17} carried out a systematic 
study on the effects of uncertainties associated with the BE on the astrochemical two-phase model 
of a dark molecular cloud. They also pointed out the importance of 
branching ratios which need experimental validation and careful implementation in astrochemical model.
They have estimated the binding energies (mentioned in the Appendix Table A1) based on the results 
presented in \cite{coll04} for the deposition of each species on a H$_2$O substrate as:
\begin{equation}
 E_{bind,X}=\frac{T_{des,X}}{T_{des,H_2O}}\times E_{bind,H_2O}
\end{equation}
where T$_{des,X}$ is the desorption temperature of species X
deposited on a H$_2$O film, T$_{des,H_2O}$ is the desorption temperature
of H$_2$O, and E$_{bind,H_2O}$ is the BE of H$_2$O. \\

Most of the neutral species, even H and H$_2$, can be physisorbed (barrier-less) onto ice 
mantles by the van der Waals force. Although the potential energy can develop a deep minimum in 
chemisorption, physisorption is more relevant for interaction with grain surface. 
In some cases, computational studies can help to derive
BEs of the interstellar species. For example, \cite{alha07} simulated adsorption 
of H atoms to ASW using classical trajectory calculations, the off-lattice kinetic Monte Carlo approach was used by
\citep{kars14} to estimate the BE of CO and CO$_2$. To simulate the adsorption
of H$_2$ with ASW and crystalline ice, classical trajectory (CT) calculations 
have been performed by \cite{horn05}. Recently, \cite{sil17} performed quantum chemical
calculations to determine BE of H and H$_2$ with astrophysical relevant surfaces.
In the absence of any experimental or theoretical data, 
as a rule of thumb, BE of an unknown species is very crudely estimated by the addition of
BEs of its reactants. But this assumption may lead to very misleading results. 
This motivates us to devise an approach to better approximate the BE for some relevant interstellar species. \\

In this paper, we present our computed BEs of $100$ interstellar and circumstellar species 
where water cluster is used as an adsorbent. Detail comparison is also made between the
calculated BEs and available experimental or theoretically obtained BE values.
High-level quantum chemical calculations are performed to 
calculate the BEs of various species. This paper is organized as follows. 
In Section 2, we discuss computational details and methodology. In Section 3, 
results are discussed in detail and finally, concluding remark is made 
in Section 4. \\

\section{Computational Details and Methodology}
The adsorption energy is usually seen as a local property arising from the electronic 
interaction between a solid support (grain surface or adsorbent) and the species 
deposited on its surface (adsorbate). We calculate the adsorption energy (BE) 
of a species on the grain surface as follows:
\begin{equation}
 E_{ads}=E_{ss}-(E_{surface}+E_{species})
\end{equation}
where E$_{ads}$ is the adsorption energy, E$_{ss}$ is the optimized energy for a species placed at a suitable distance 
from the grain surface, E$_{surface}$ and E$_{species}$ are the optimized energies of the grain surface and species respectively.
According to various studies, around dense cloud regions, water (H$_2$O) is the major ($\sim$70\% by mass)
constituent of a grain mantle \citep{kean01,das10,das11,das16}, and thus the incoming gas species may be directly adsorbed onto 
the water ice. So, a knowledge of the BE of the adsorbed species with water ice is essential
to build a realistic astrochemical model which studies the composition of the interstellar grain mantle.
For our investigation on the BEs, we use the most stable configurations of 
water monomer, c-trimer, c-tetramer, c-pentamer, and c-hexamer (chair) \citep{ohno05} respectively 
as the adsorbents (Fig. 1). \\

\begin{figure}
\centering
\includegraphics[width=\textwidth]{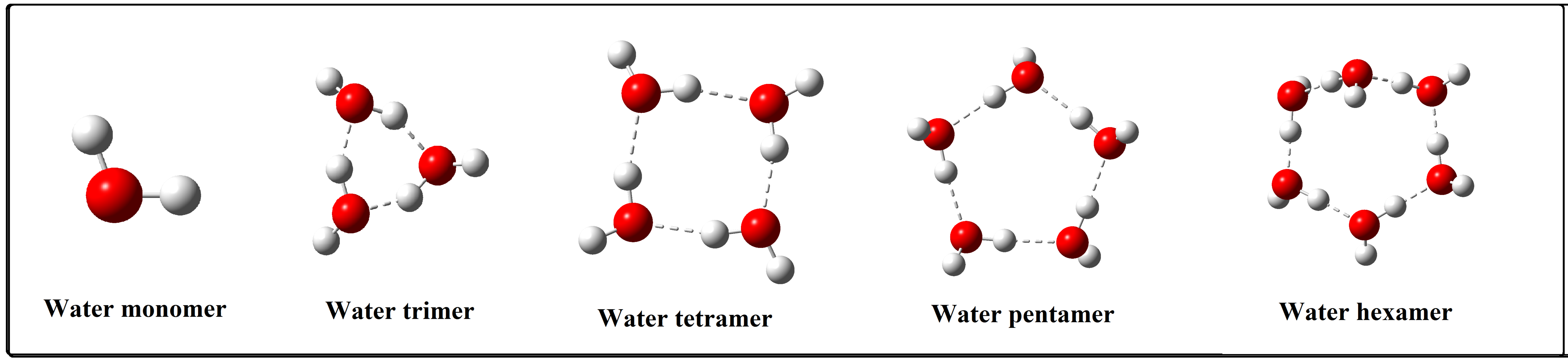}
\caption{Configurations of water molecule(s) which are used in the present work.}
\end{figure}

We carry out all the calculations by using Gaussian 09 suite of programs developed by \cite{fris13}.
Second order M$\o$ller-Plesset (MP2) method with aug-cc-pVDZ basis set is  mainly used in computing the optimized
energy of all the species and complexes. 
A prefix `aug' is used to indicate the addition of diffuse function and cc-pVDZ,  cc-pVTZ are
the Dunning’s correlation consistent basis set \citep{dunn89} having the double  and triple zeta function respectively. 
CCSD(T) (Coupled Cluster single-double and perturbative triple)
method with aug-cc-pVTZ basis set is also used to calculate the single point energy by taking the optimized structure obtained with MP2
method for some 
tetramer and hexamer (water cluster) to check the dependency of our computed BE values on the implemented method and
basis set.
Fully optimized ground state structure is verified to be a stationary point (having non-negative 
frequency) by harmonic vibrational frequency analysis and most of the calculations are performed without 
zero point energy (ZPE) and without basis set superposition error (BSSE) corrections.
A study have also been carried 
out by including ZPE and excluding BSSE (using the Counterpoise method) to check their influence on calculated BE values.\\

\begin{sidewaystable}
\scriptsize{
\caption{Calculated BEs Along with Some Frequently Used BEs for Astrochemical Modeling.}
\hskip -4.6cm
\begin{tabular}{||c|c|c|c|c|c|c|c|c|c|c||}
\hline
\hline
{\bf Sl.} & {\bf Species} & \multicolumn{5}{|c|}{\bf Calculated BE in Kelvin on different water clusters using MP2/aug-cc-pVDZ} 
&{\bf Experimental} & \multicolumn{2}{|c|}{\bf \underline{BE from \cite{wake17}}}& {\bf BE in Kelvin} \\
\cline{3-7}
{\bf No.}&&{\bf Monomer}&{\bf Trimer}&{\bf Tetramer}&{\bf Pentamer}&{\bf Hexamer}&{\bf values of BE} 
&\underline{{\bf in Kelvin}} &\underline{{\bf in Kelvin}}&{\bf from}\\
& & \includegraphics[width=1.3cm, height=0.8cm]{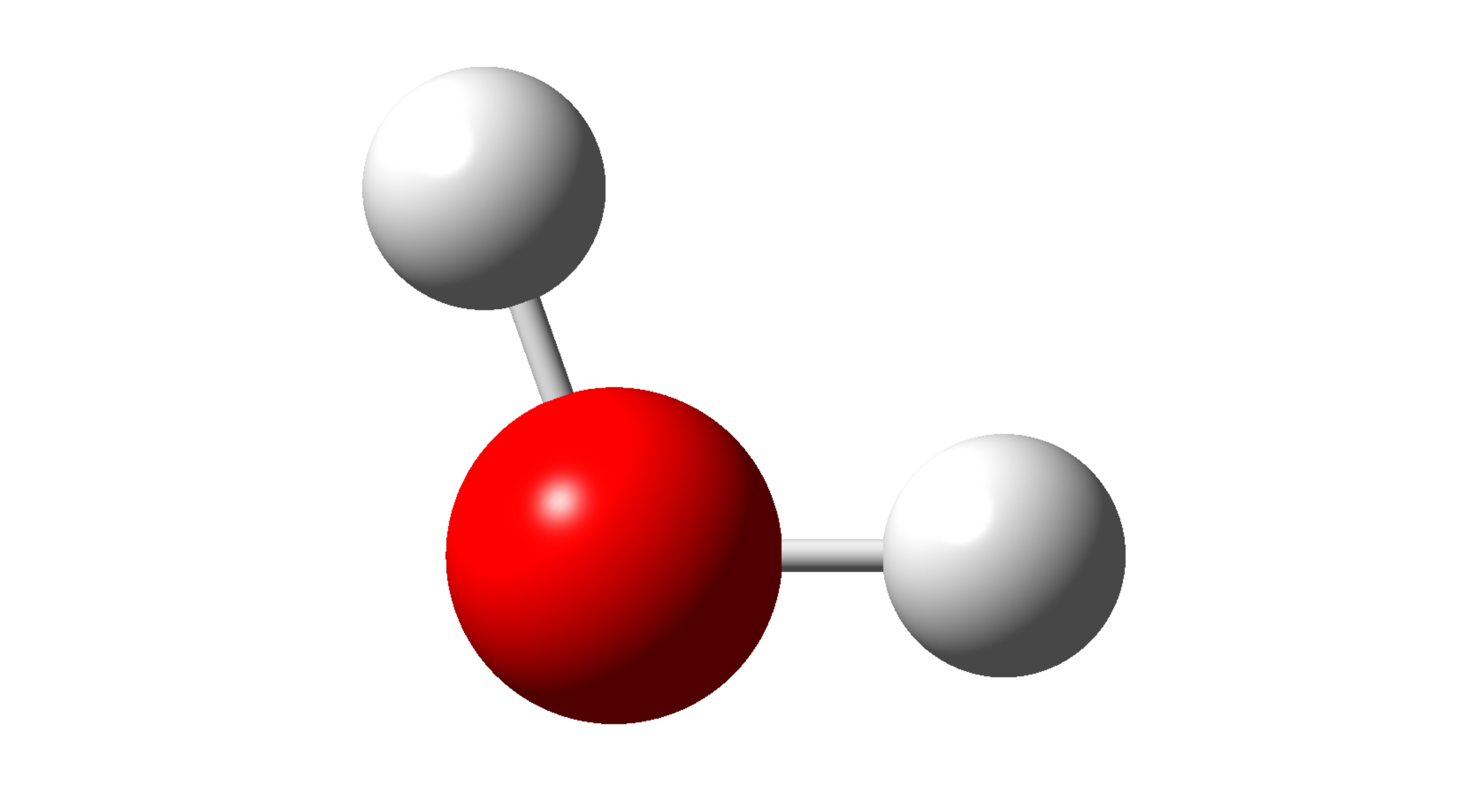} &
\includegraphics[width=1.5cm, height=1cm]{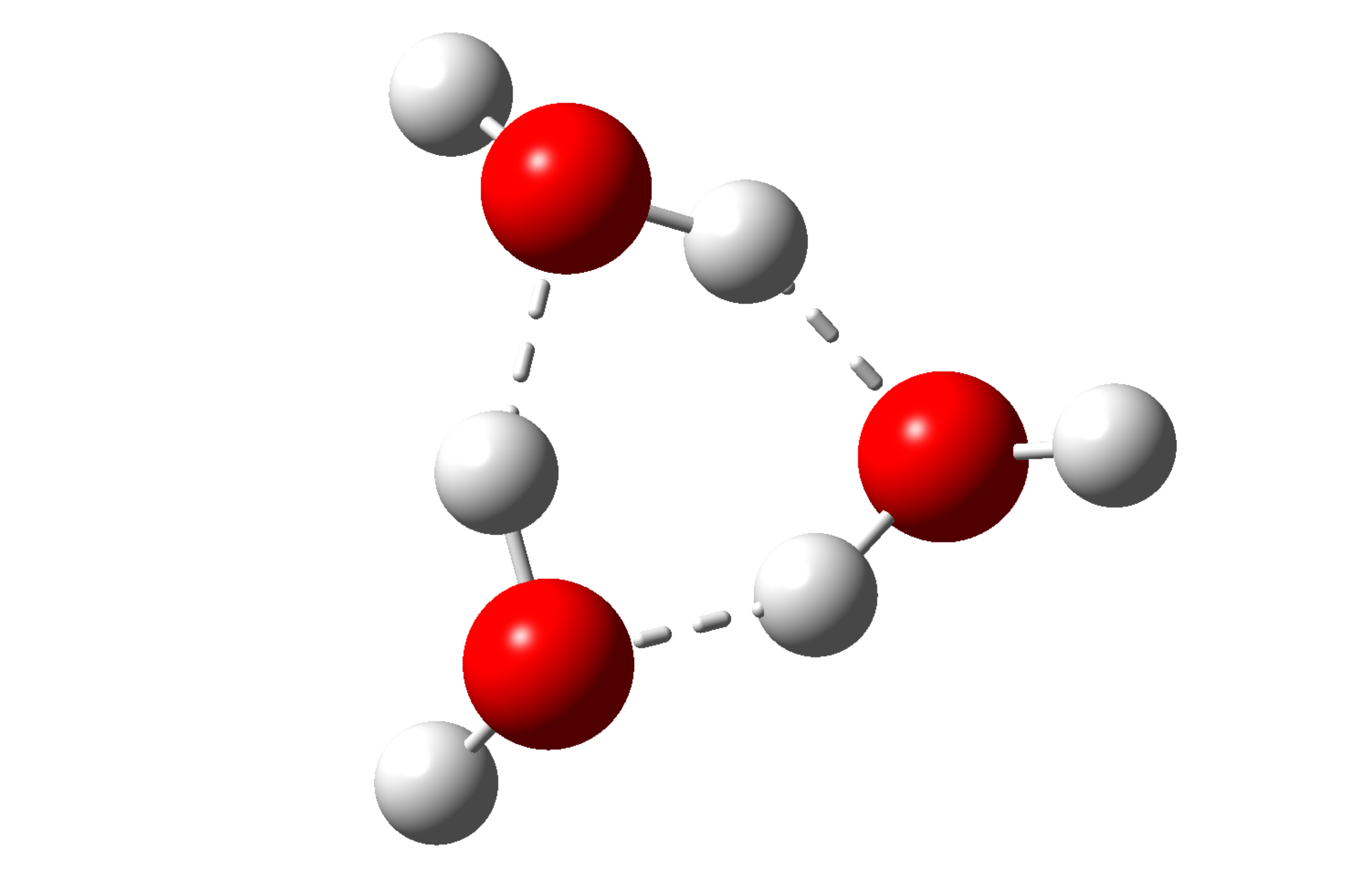} & \includegraphics[width=1.7cm, height=1.2cm]{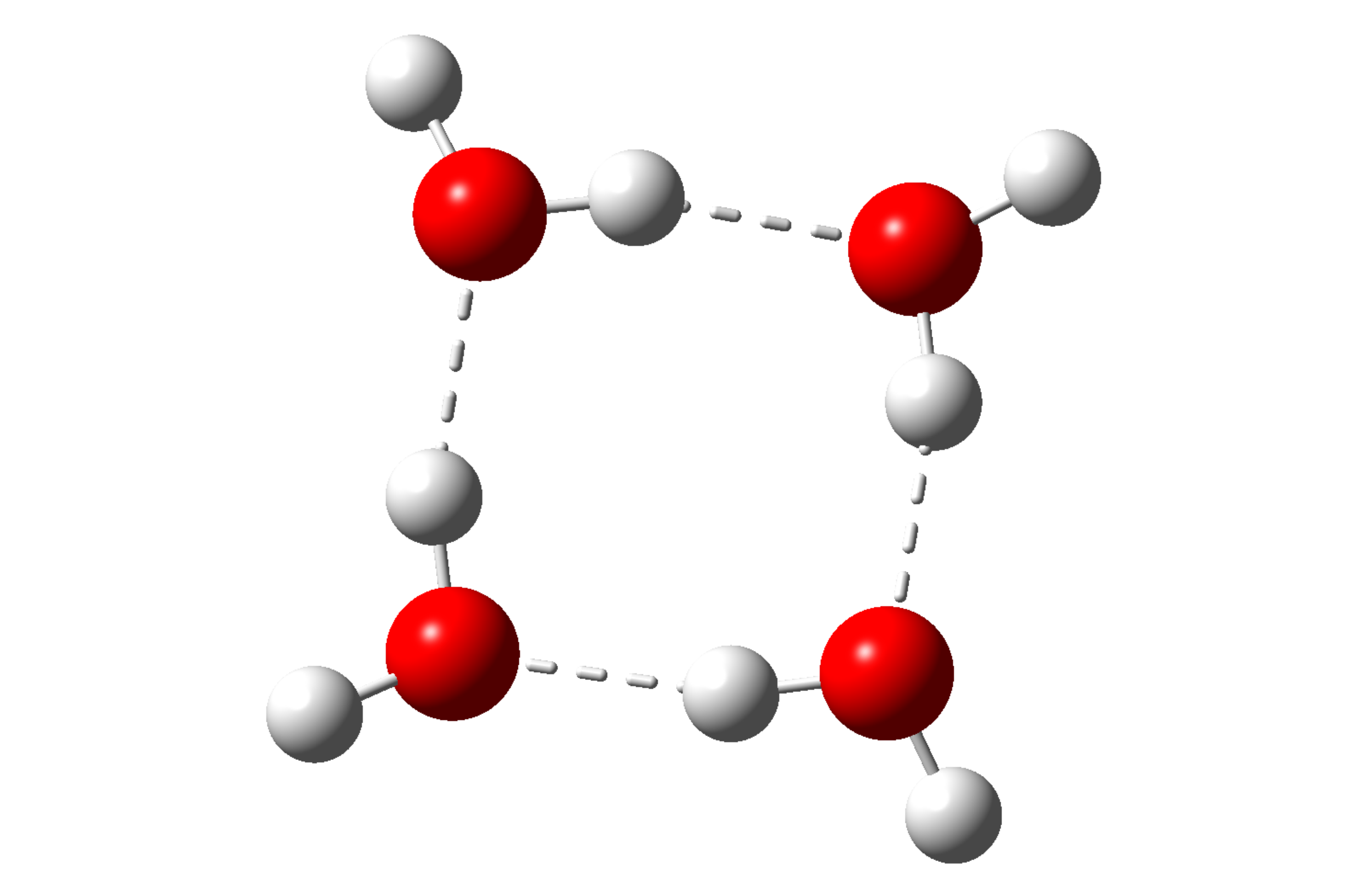}
& \includegraphics[width=1.7cm, height=1.2cm]{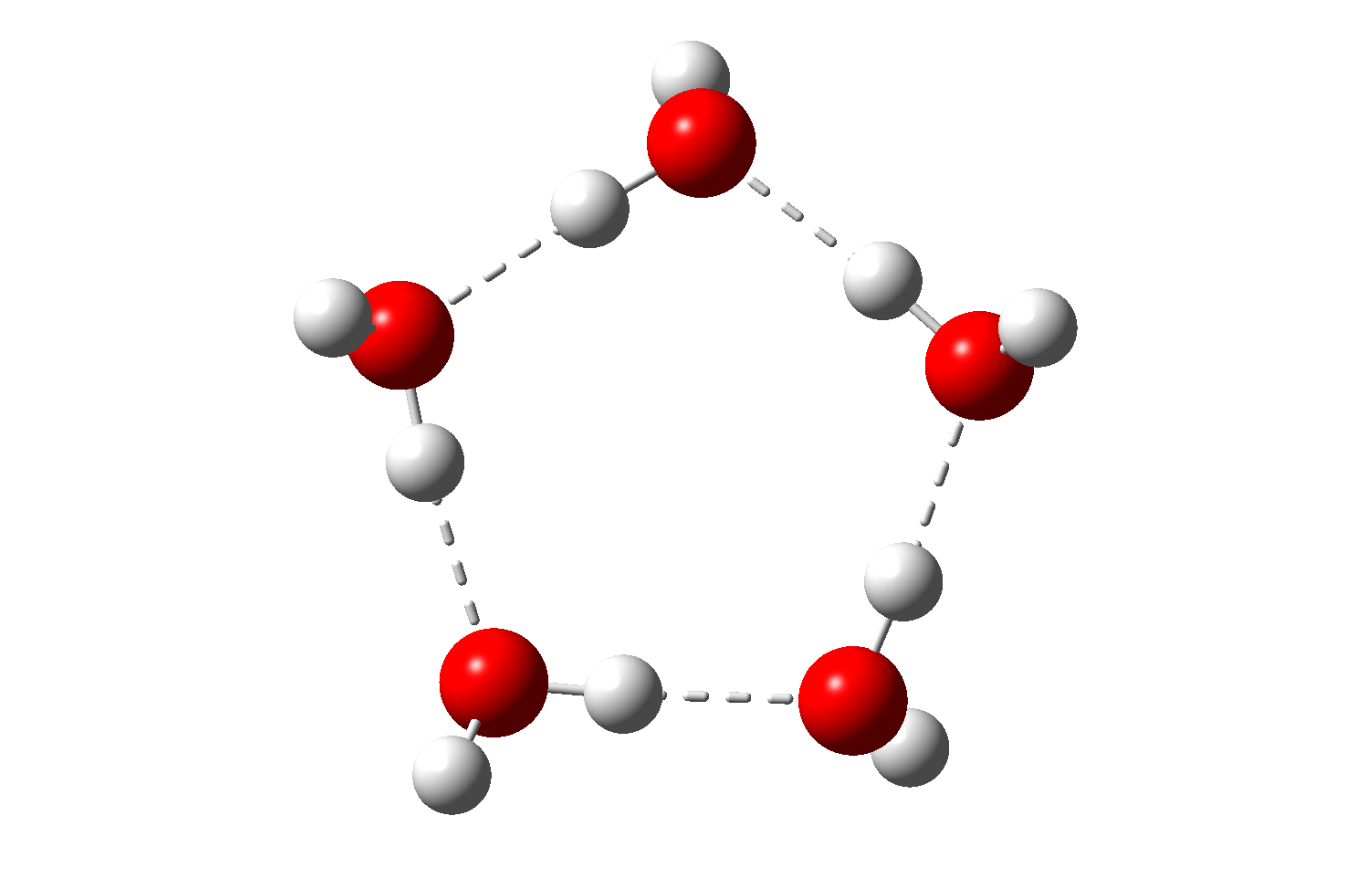} & \includegraphics[width=1.7cm, height=1.2cm]{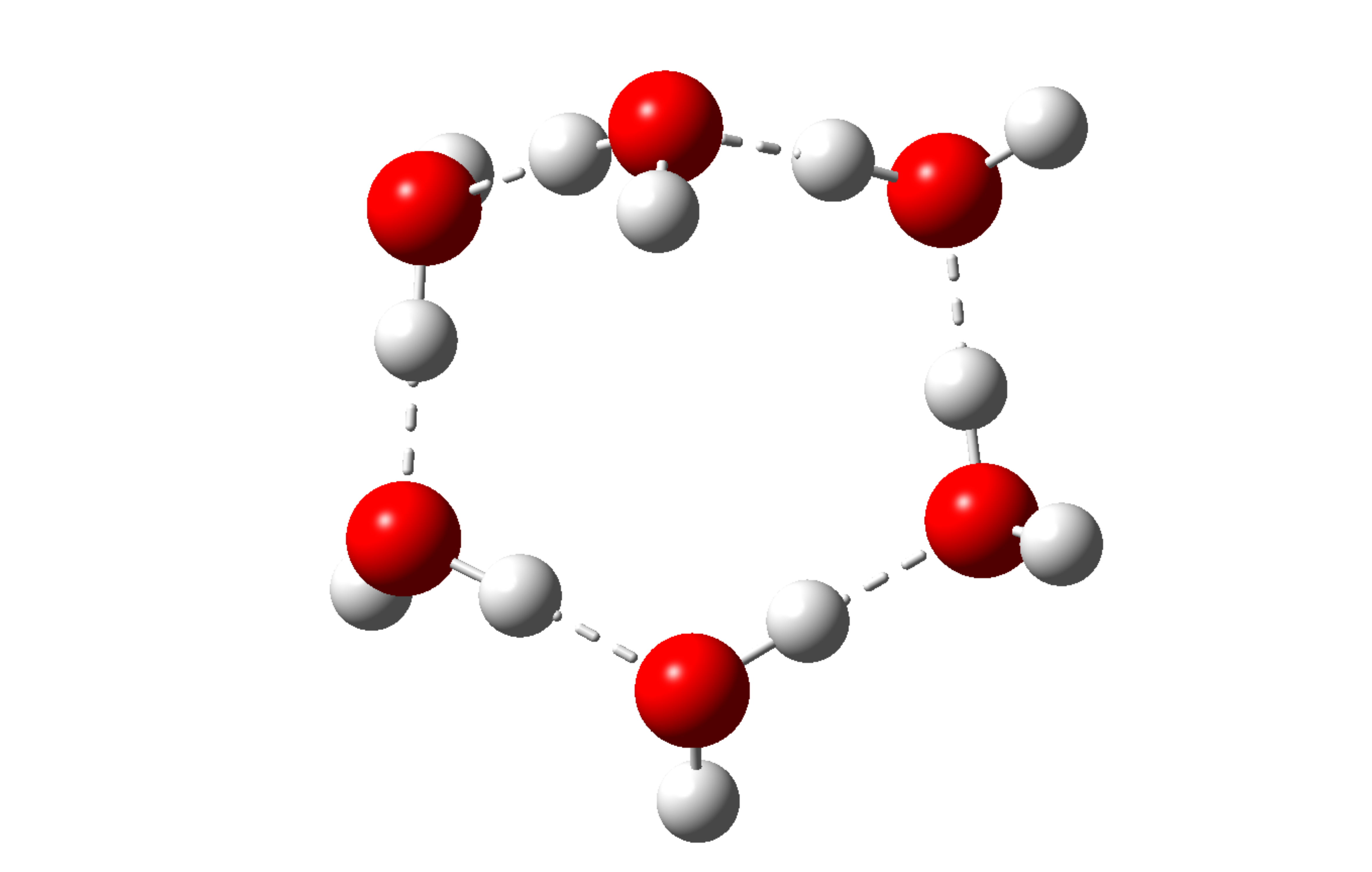} & 
{\bf in Kelvin} & {\bf using DFT} & {\bf using MP2}& {\bf UMIST database$^k$} \\
&&&&&&&& {\bf  M06-2X} & {\bf aug-cc-pVTZ}&  \\
\hline
\hline
1. & OCS & 1139 (-53.1 \%) &  1905 (-21.6 \%) & 1571 (-35.3 \%) & 2014 (-17.1 \%)& 1808 (-25.5 \%) & 2430 $\pm$ 24$^a$  & 2100 (13.5 \%) & --- & 2888 (18.8 \%) \\
2. & HCl & 3116 (-39.7 \%) &3545 (-31.4 \%) & 3924 (-24.1 \%) & 4099 (-20.7 \%) & 4104 (-20.6 \%) & 5172$^b$  & 4800 (7.1 \%) & 5000 (-3.3 \%)  & 900 (-82.5 \%) \\
3. & CH$_3$CN & 2676 (-42.8 \%) & 4108 (-12.2 \%) & 2838 (-39.3 \%) & 2820 (-39.7 \%) & 3786 (-19.1 \%) & 4680$^c$ & 4300 (8.1 \%) & 4800 (2.5 \%) &  4680 (0.00 \%) \\
4. & H$_2$O$_2$ & 3838 (-36.0 \%) &  3942 (-34.3 \%) & 3928 (-34.5 \%) & 5288 (-11.8 \%) & 5286 (-11.9 \%) & 6000$^d$  & 6800 (-13.3 \%) & 6100 (1.6 \%) & 5700 (-5.0 \%)\\
5. & CH$_3$OH & 3124 (-37.5 \%) &  3924 (-21.5 \%)& 4368 (-12.6 \%) & 4607 (-7.8 \%) & 4511 (-9.7 \%) & 5000$^e$  & 4800 (4.0 \%) & 4850 (-3.0 \%)& 4930 (-1.4 \%) \\
6. & NH$_3$ & 3501 (-36.6 \%) & 3745 (-32.3 \%) & 3825 (-30.8 \%) & 3751 (-32.1 \%) & 5163 (-6.6 \%) & 5530$^c$ & 5600 (-1.2 \%) & 5500 (-0.5 \%) &  5534 (0.6 \%) \\
7. & CO & 595 (-54.2 \%) &  664 (-48.9 \%) & 1263 (-2.8 \%) & 1320 (1.5 \%) & 1292 (-0.6 \%) & 1300$^{g}$ & 1300 (0.0 \%) & 1100 (-15.3 \%) & 1150 (-11.5 \%) \\
8. & C$_2$H$_2$ & 1532 (-40.7 \%) & 2581 (-0.2 \%) & 2593 (0.2 \%) & 2633 (1.7 \%) & 2640 (2.0 \%) & 2587$^c$ & 2600 (-0.5 \%) & 2700 (4.3 \%) &  2587 (0.0 \%) \\
9. & CO$_2$ & 1506 (-34.5 \%) & 1935 (-15.9 \%) & 2293 (-0.3 \%) & 2287 (-0.5 \%) & 2352 (2.2 \%) & 2300$^g$ & 3100 (34.7 \%) & 2600 (13.0 \%) &  2990 (30.0 \%) \\
10. & N$_2$ & 793 (-27.9 \%) & 884 (-19.6 \%)& 900 (-18.1 \%) & 893 (-18.8 \%) & 1161 (5.5 \%) & 1100$^e$ & 1100 (0.0 \%) & 1400 (27.2 \%) & 790 (-28.1 \%) \\
11. & NO & 886 (-50.7 \%) & 568 (-68.4 \%)& 1265 (-29.7 \%) & 1300 (-27.7 \%) & 1988 (10.4 \%) & 1800$^e$ & 1600 (11.1 \%) & 1700 (-5.5 \%) &  1600 (-11.1 \%) \\
12. & O$_2$ & 391 (-67.4 \%) &  382 (-68.2 \%) & 940 (-21.6 \%) & 1116 (-7.0 \%) & 1352 (12.6 \%) & 1200$^g$ & 1000 (16.6 \%) & 900 (-25.0 \%) & 1000 (-16.6 \%) \\
13. & H$_2$S & 1727 (-37.0 \%) & 2849 (3.9 \%) & 2556 (-6.8 \%) & 2396 (-12.6 \%) & 3232 (17.8 \%) & 2743$^{c,k}$ & 2700 (1.5 \%) & 2550 (-7.0 \%) &  2743 (0.0 \%) \\
14. & CH$_3$CCH & 2266 (-9.3 \%) &  2580 (3.2 \%)& 2342 (-6.3 \%) & 2673 (6.9 \%) & 3153 (26.1 \%) & 2500 $\pm$ 40$^h$ & 3800 (-52.0 \%) & 3800 (52.0 \%) & 2470 (-1.2 \%) \\
15. & HNCO & 2046 (-47.5 \%) &  3973 (1.9 \%) & 3922 (0.5 \%) & 4097 (5.0 \%) & 5554 (42.4 \%) & 3900$^i$  & 4850 (-24.3 \%) & --- &  2850 (-26.9 \%) \\
16. & CH$_4$ & 469 (-51.2 \%) & 1066 (10.7 \%) & 1327 (37.7 \%) & 1366 (41.8 \%) & 1491 (54.8 \%) & 963$^j$ & 800 (16.9 \%) & 1000 (3.8 \%) & 1090 (13.1 \%) \\
\hline
\hline
\multicolumn{2}{||c|}{\bf  Average absolute deviation}& $\pm$ 41.6 \%& $\pm$ 24.6 \%
&  $\pm$ 18.8 \% & $\pm$ 15.8 \% & $\pm$ 16.7 \% &  & $\pm$ 12.8 \%
& $\pm$ 11.7 \% &  $\pm$ 15.4 \%\\
\hline
\multicolumn{2}{||c|}{\bf Fractional RMS deviation} &  0.435 &  0.324 &  0.236
&  0.205 &0.221 &  &  0.189 &  0.182 & 0.254\\
\hline
\hline
\end{tabular}} \\
\\
{\bf Notes.} Percentage deviation from experimental BE values (column 8) are shown in parentheses for columns 3, 4, 5, 6, 7, 9, 10 and 11. \\
$^a$ \cite{ward12}, $^b$ \cite{olan11}, $^c$ \cite{coll04}, $^d$ \cite{duli13}, $^e$ \cite{wake17},
$^g$ \cite{mini16}, $^h$ \cite{kimb14}, $^i$ \cite{nobl15}, $^j$ \cite{raut07},
$^k$ UMIST database (\url{http://udfa.ajmarkwick.net}).
\end{sidewaystable}

\begin{table}
\scriptsize{
\caption{Comparison of Calculated BEs Using Water Monomer (adsorbent) with Experimentally used BEs.}
\hskip -3.8cm
\begin{tabular}{||c|c|c|c|c|c|c|c||}
\hline
\hline
{\bf Sl.} & {\bf Species} & \multicolumn{5}{|c|}{\bf BE in Kelvin using different methods and basis sets 
including (+) or excluding (+) ZPE and BSSE} & {\bf Experimental} \\
\cline{3-7}
{\bf No.}&&{\bf MP2/aug-cc-pVDZ}&{\bf MP2/aug-cc-pVDZ}&{\bf MP2/aug-cc-pVDZ}&{\bf MP2/aug-cc-pVTZ} & {\bf CCSD(T)/aug-cc-pVTZ} & {\bf values of BE} \\

&& {\bf - ZPE and - BSSE} & {\bf + ZPE but - BSSE} & {\bf - ZPE but + BSSE} & {\bf - ZPE and - BSSE} & {\bf - ZPE and - BSSE} & {\bf in Kelvin} \\

\hline
\hline
1. & OCS & 1139 (-53.1 \%) & 683 (-71.9 \%) & 803 (-66.9 \%) &  1074 (-55.8 \%) & 1086 (-55.3 \%) & 2430 $\pm$ 24$^a$ \\
2. & HCl & 3116 (-39.7 \%) & 2113 (-59.1 \%) & 2627 (-49.2 \%) &  2975 (-42.5 \%) & 2777 (-46.3 \%) & 5172$^b$ \\
3. & CH$_3$CN & 2676 (-42.8 \%) & 1970 (-57.9 \%) & 2242 (-52.1 \%) & 2676 (-42.8 \%) & 2635 (-43.7 \%) & 4680$^c$ \\
4. & H$_2$O$_2$ & 3838 (-36.0 \%) & 2647 (-55.9 \%) & 3204 (-46.6 \%) & 3775 (-37.1 \%) & 3802 (-36.6 \%) & 6000$^d$  \\
5. & CH$_3$OH & 3124 (-37.5 \%) & 2149 (-57.0 \%) & 2586 (-48.3 \%) & 3021 (-39.6 \%) & 2988 (-40.2 \%) & 5000$^e$ \\
6. & NH$_3$ & 3501 (-36.6 \%) & 2368 (-57.2 \%) & 2941 (-46.8 \%) & 3375 (-39.0 \%) & 3332 (-39.7 \%) & 5530$^c$  \\
7. & CO & 595 (-54.2 \%) & 236 (-81.8 \%) & 349 (-73.2 \%) & 565 (-56.5 \%) & 662 (-49.1 \%) & 1300$^g$  \\
8. & C$_2$H$_2$ & 1532 (-40.7 \%) & 950 (-63.3 \%) & 1111 (-57.1 \%) & 1519 (-41.3 \%) & 1444 (-44.2 \%) & 2587$^c$   \\
9. & CO$_2$ & 1506 (-34.5 \%) & 1109 (-51.8 \%) & 1159 (-49.6 \%) & 1417 (-38.4 \%) & 1511 (-34.3 \%) & 2300$^g$  \\
10. & N$_2$ & 793 (-27.9 \%) & 340 (-69.1 \%) & 534 (-51.4 \%) & 791 (-28.1 \%) & 759 (-31.0 \%) & 1100$^e$  \\
11. & NO & 886 (-50.7 \%) & 353 (-80.4 \%) & 249 (-86.2 \%) &  876 (-51.3 \%) & 780 (-56.7 \%) & 1800$^e$   \\
12. & O$_2$ & 391 (-67.4 \%) & 258 (-78.5 \%) & 191 (-84.1 \%) & 385 (-67.9 \%) & 419 (-65.1 \%) & 1200$^g$   \\
13. & H$_2$S & 1727 (-37.0 \%) & 971 (-64.6 \%) & 1305 (-52.4 \%) & 1662 (-39.4 \%) & 1598 (-41.7 \%) & 2743$^{c,k}$  \\
14. & CH$_3$CCH & 2266 (-9.3 \%) & 1548 (-38.1 \%) & 1675 (-33.0 \%) & 2175 (-13.0 \%) & 2083 (-16.7 \%) & 2500 $\pm$ 40$^h$ \\
15. & HNCO & 2046 (-47.5 \%) & 1376 (-64.7 \%) & 1644 (-57.8 \%) & 3260 (-16.4 \%) & 2058 (-47.2 \%) & 3900$^i$  \\
16. & CH$_4$ & 469 (-51.2 \%) & 145 (-84.9 \%) & 265 (-72.5 \%) & 374 (-61.2 \%) & 401 (-58.4 \%) & 963$^j$ \\
\hline
\hline
\multicolumn{2}{||c|}{\bf Average abs deviation}  & $\pm$ 41.6 \% & $\pm$ 64.7 \% & $\pm$ 58 \% & $\pm$ 41.9 \% & $\pm$ 44.1 \% & --- \\
\hline
\multicolumn{2}{||c|}{\bf Frac. RMS deviation}  & 0.435 & 0.659 & 0.597 & 0.443 & 0.456 & --- \\
\hline
\hline
\end{tabular}} \\
\\
{\bf Notes.} Percentage deviation from experimental BE values (column 8) are shown in parentheses for columns 3, 4, 5, 6, and 7. \\
$^a$ \cite{ward12}, $^b$ \cite{olan11}, $^c$ \cite{coll04}, $^d$ \cite{duli13}, $^e$ \cite{wake17},
$^g$ \cite{mini16}, $^h$ \cite{kimb14}, $^i$ \cite{nobl15}, $^j$ \cite{raut07},
$^k$ UMIST database (\url{http://udfa.ajmarkwick.net}).
\end{table}

\begin{table}
\centering{
\scriptsize{
\caption{Calculated BEs using water hexamer (adsorbent) to check the effect of basis set superposition error (BSSE) using 
counterpoise method.}
\hskip -2.7cm
\begin{tabular}{||c|c|c|c||}
\hline
\hline
{\bf Sl.} & {\bf Species} & \multicolumn{2}{|c||}{\bf BE in Kelvin using MP2/aug-cc-pVDZ method} \\
\cline{3-4}
{\bf No.}&&{\bf Values without BSSE correction} & {\bf BSSE corrected values} \\
\hline
\hline
1. & OCS & 1808 (-25.5 \%) & 1294 (-46.7 \%) \\
2. & HCl & 4104 (-20.6 \%) & 3777 (-26.9 \%) \\
3. & $\rm{CH_3CN}$ & 3786 (-19.1 \%) & 2194 (-53.1 \%) \\
4. & $\rm{H_2O_2}$ & 5286 (-11.9 \%) & 4161 (-30.6 \%) \\
5. & $\rm{CH_3OH}$ & 4511 (-9.7 \%) & 3550 (-29.0 \%) \\
6. & $\rm{NH_3}$ & 5163 (-6.6 \%) & 3082 (-44.2 \%) \\
7. & CO & 1292 (-0.6 \%) & 840 (-35.3 \%) \\
8. & $\rm{C_2H_2}$ & 2640 (2.0 \%) & 1890 (-26.9 \%) \\
9. & $\rm{CO_2}$ & 2352 (2.2 \%) & 1624 (-29.3 \%) \\
10. & N$_2$ & 1161 (5.5 \%) & 568 (-48.3 \%) \\
11. & NO & 1988 (10.4 \%) & 911 (-49.3 \%) \\
12. & O$_2$ & 1352 (12.6 \%) & 519 (-56.7 \%) \\
13. & $\rm{H_2S}$ & 3232 (17.8 \%) & 1954 (-28.8 \%) \\
14. & $\rm{CH_3CCH}$ & 3153 (26.1 \%) & 1382 (-44.7 \%) \\
15. & HNCO & 5554 (42.4 \%) & 5017 (28.6 \%) \\
16. & $\rm{CH_4}$ & 1491 (54.8 \%) & 653 (-32.1 \%) \\
\hline
\hline
\multicolumn{2}{||c|}{\bf Average absolute deviation} & $\pm$ 16.7 \% & $\pm$ 34.6 \% \\
\hline
\multicolumn{2}{||c|}{\bf Fractional RMS deviation} & 0.221 & 0.395 \\
\hline
\hline
\end{tabular}}} \\
{\bf Notes.} Percentage deviation from experimental BE values are shown in parentheses for columns 3 and 4.
\end{table}

\begin{table}
\centering{
\scriptsize{
\caption{Calculated BEs using water tetramer (adsorbent) to check the effect of higher order quantum chemical method.}
\hskip -2.7cm
\begin{tabular}{||c|c|c|c||}
\hline
\hline
{\bf Sl.} & {\bf Species} & \multicolumn{2}{|c||}{\bf BE in Kelvin using water tetramer} \\
\cline{3-4}
{\bf No.}&&{\bf MP2/aug-cc-pVDZ}&{\bf CCSD(T)/aug-cc-pVTZ} \\
\hline
\hline
1. & N & 269 & 273 \\
2. & O & 1002 & 1024 \\
3. & O$_2$ & 940 & 853 \\
4. & H$_2$O & 2670 & 2632 \\
5. & CO & 1263 & 1196 \\
6. & N$_2$ & 900 & 854 \\
\hline
\hline
\end{tabular}}}
\end{table}

\section{Results and Discussions over Calculated Binding Energy Values}
In this Section, results of high-level quantum chemical calculations are presented and 
discussed in detail. 
\cite{wake17} selected $16$ stable species to compute the BEs by considering a water
monomer. Here, we employ similar methodology and carry out calculations by increasing the 
cluster size to check its consequences. Five different sets of adsorbent are used to see the effects of
cluster size on the computed BEs. With the increase in cluster size, we found an increasing trend of BEs for most 
of the species considered here. Most interestingly the
calculated BEs by considering water c-pentamer and c-hexamer (chair) configuration seems to be 
closer to the experimentally obtained values as compared to that of the calculated BEs with 
the water monomer, c-trimer, or c-tetramer configurations.
Binding energies of these $16$ stable species with various water cluster configurations are given in Table 1.
For the sake of comparison, in Table 1, we also show 
the experimentally obtained values, estimated values from \cite{wake17},
and BEs from UMIST (\url{http://udfa.ajmarkwick.net}) database. \\

\begin{figure}
\centering
\includegraphics[width=\textwidth]{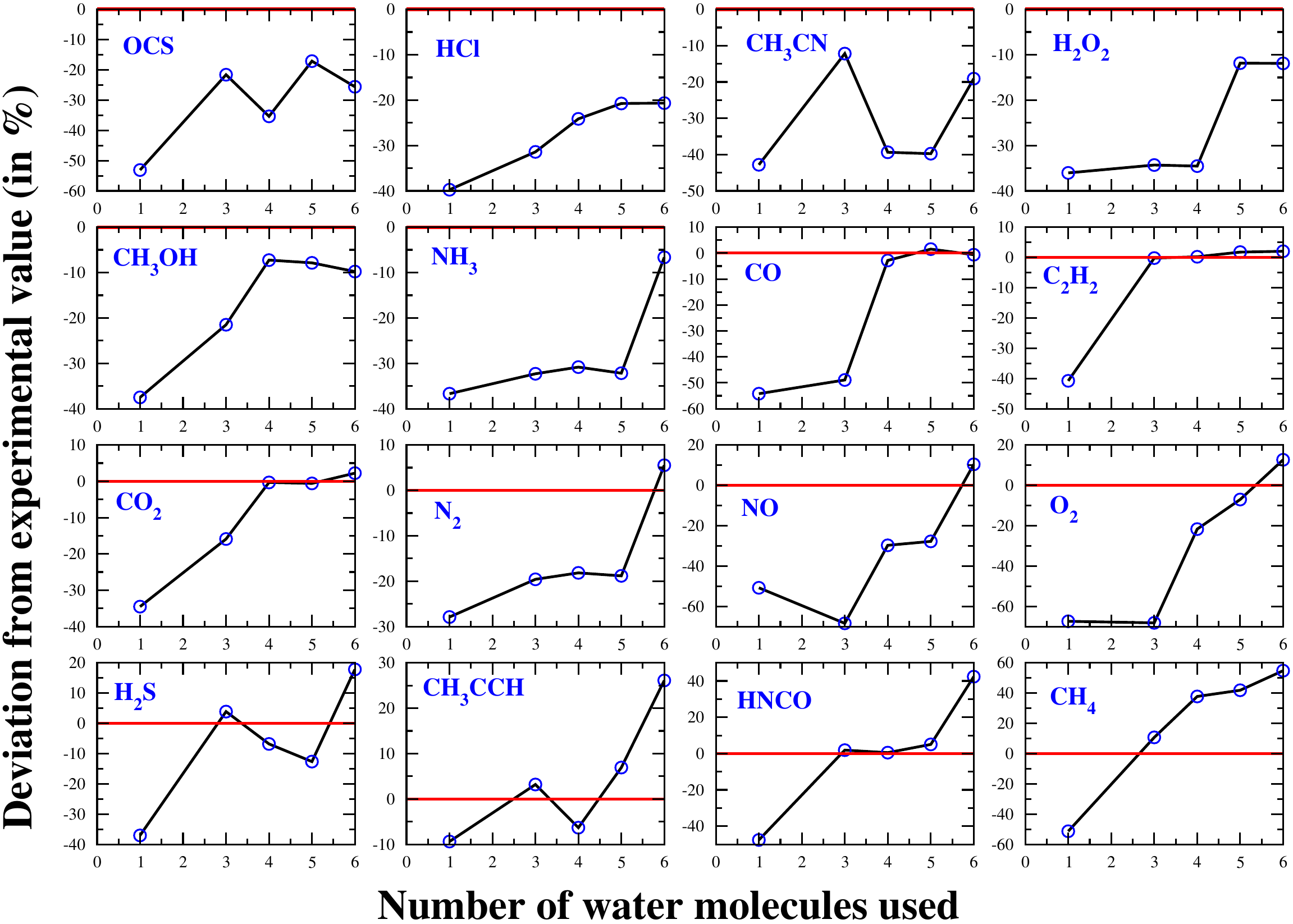}
\caption{Percentage deviation of BEs of $16$ stable species with increasing number of water clusters acting as the 
grain surface.}
\end{figure}

For most of the cases, Fig. 2 shows an increasing trend of BEs with the increase in the size of 
the water cluster as an adsorbent.
Figure 2 is subdivided into $16$ blocks for $16$  different species.
Along the X-axis, we show the size of water cluster and along Y-axis, we show the 
percentage deviation from the experimentally obtained values. The horizontal red line in each block
denotes the zero deviation from experimental values.
The result clearly shows that for most of the cases the percentage deviation between
experiment and theory is comparatively higher (underestimated) when the water monomer is used 
and it is lower when c-pentamer or c-hexamer (chair) configuration of water is
used as an absorbent. All the percentage deviations of each species are shown in parentheses of Table 1
and Table 2 as well. In order to check the overall comparison between our calculations and experimentally obtained values,
we calculate the average absolute percentage deviation and root mean square fractional deviation of these species
and these are also presented in Table 1 and Table 2. Figure 3 depicts the percentage deviation from experimental BE values of each
species for water hexamer. It is striking to note that (in Table 1) while we sequentially increase the cluster
size by considering water monomer, c-trimer, c-tetramer, c-pentamer and c-hexamer (chair) as a substrate, 
the average percentage deviations are found to be $ \pm 41.6 \%$, $ \pm 24.6 \%$, $ \pm 18.8 \%$, $ \pm 15.8 \%$, 
and  $ \pm 16.7 \%$ respectively and fractional root mean square deviations are found to be 0.435, 0.324, 0.236, 0.205, and 0.221
respectively. These average absolute percentage deviations and fractional R.M.S. deviations are also shown in Fig. 4(a) and 
4(b) respectively for a better understanding. \\

In Table 1, we also present the average percentage deviation of calculated  BE values from \cite{wake17}. 
We see that on an average, the predicted/scaled values of \cite{wake17} deviate the experimental
values by $ \pm 12.8 \%$ while M06-2X method with aug-cc-pVTZ basis set was used and deviate the value by 
$ \pm 11.7 \%$ while MP2 method with aug-cc-pVTZ basis set was used. It is also shown that with respect to the experimental
values noted in Table 1, BE values in UMIST database deviates the experimental values by $ \pm 15.4 \%$.
Our calculation with  c-pentamer and c-hexamer (chair) configuration are producing results (no fitting is
required as in \cite{wake17}) which on an average  deviate the results $\pm 15.8 \%$ and $\pm 16.7 \%$ 
respectively. Since, the calculations performed with  c-pentamer and c-hexamer (chair) 
configurations can roughly estimate the experimental values, in the absence of the experimental values it is suggested to use
these configurations. In Table 2, we have shown a comparison between the results obtained by
considering BSSE corrections using counterpoise method (column 5 of Table 2) and without BSSE corrections (column 3 of Table 2)
with water monomer. Considering water c-hexamer (chair) structure, same comparison is performed in Table 3.
BSSE corrected BE values are found to be lower than that of without BSSE correction which imply that
the basis set leads to significant BSSE. We have also shown the comparison between the results obtained by including ZPE 
(column 4 of Table 2) and without inclusion of ZPE (column 3 of Table 2) in case of only water monomer.
It is to be noted that results obtained without inclusion of ZPE and BSSE corrections are more closer to the experimental values. 
Also we have tested our BE calculations with water monomer by using higher level method and larger basis set (CCSD(T)/aug-cc-pVTZ)
by single point energy calculation (column 7 of Table 2). We show in Table 4 the BE calculations of some selected
species (N, O, O$_2$, H$_2$O, CO and N$_2$) with water tetramer using higher method and basis set (CCSD(T)/aug-cc-pVTZ). 
We found a minor change with these considerations except the effect of ZPE
though in \cite{wake17} case, the ZPE effect is small because the ZPE is roughly proportional to the BE and after
inclusion of the ZPE the parameters for the fits are not the same but the result of the fits show similar deviation. 
However, we noticed that among all the methods or basis sets used in our computation,
MP2 method with aug-cc-pVDZ basis set (without considering ZPE and BSSE corrections)
stands comparatively best with respect to the experimental values. The fact that a low level of theory (small basis set, no BSSE correction, no ZPE) results is in better agreement with experiment 
than a higher level is likely due to the fact that the various approximations at low level compensate each other. This is probably
a coincidence, but is useful at it makes it easier to process a large number of systems.\\\\

\begin{figure}
\centering
\includegraphics[width=\textwidth]{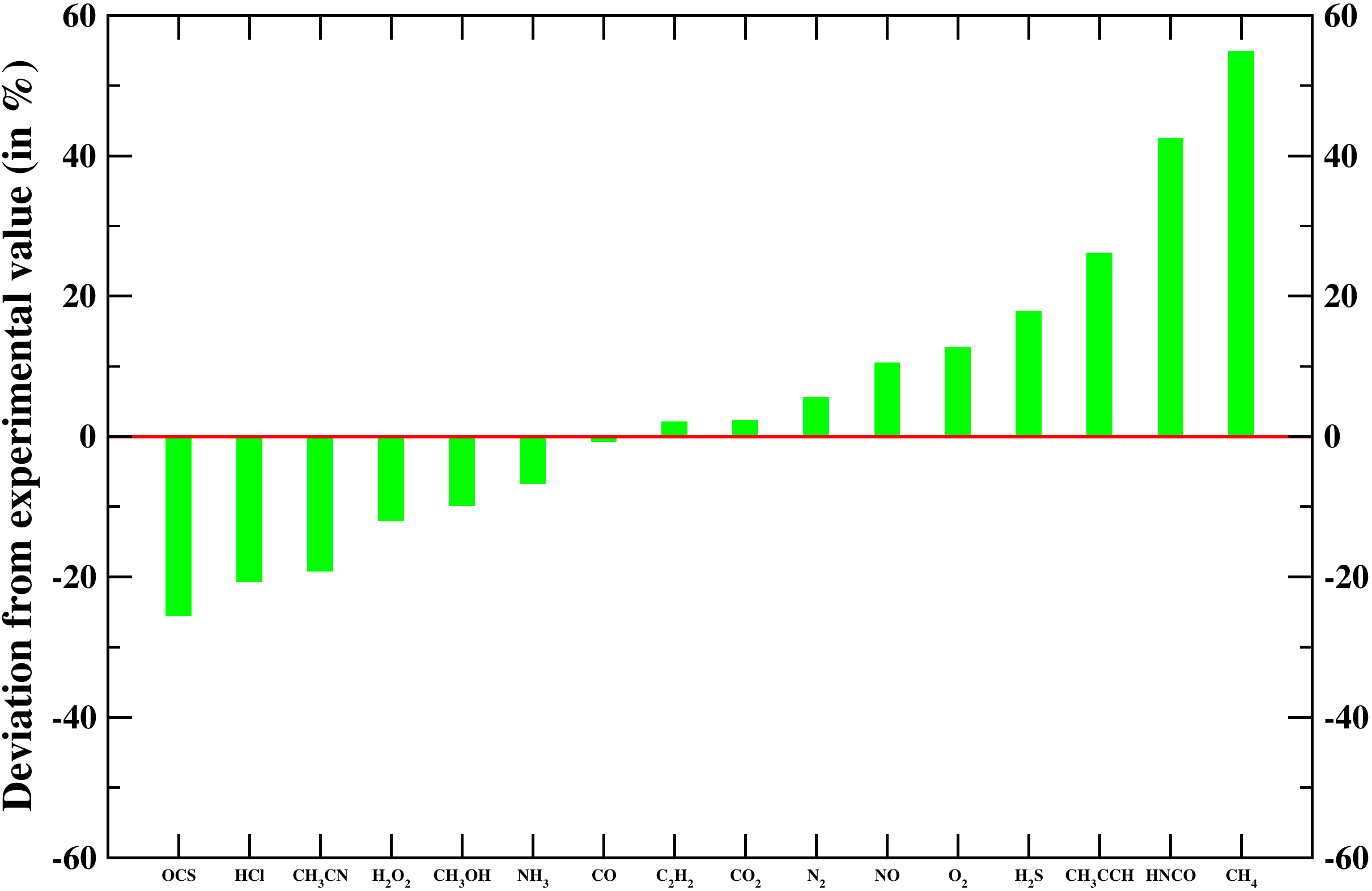}
\caption{Percentage deviation from experimental BE values of 16 stable species using water c-hexamer (chair) cluster.}
\end{figure}

\begin{figure}
\centering
\includegraphics[width=\textwidth]{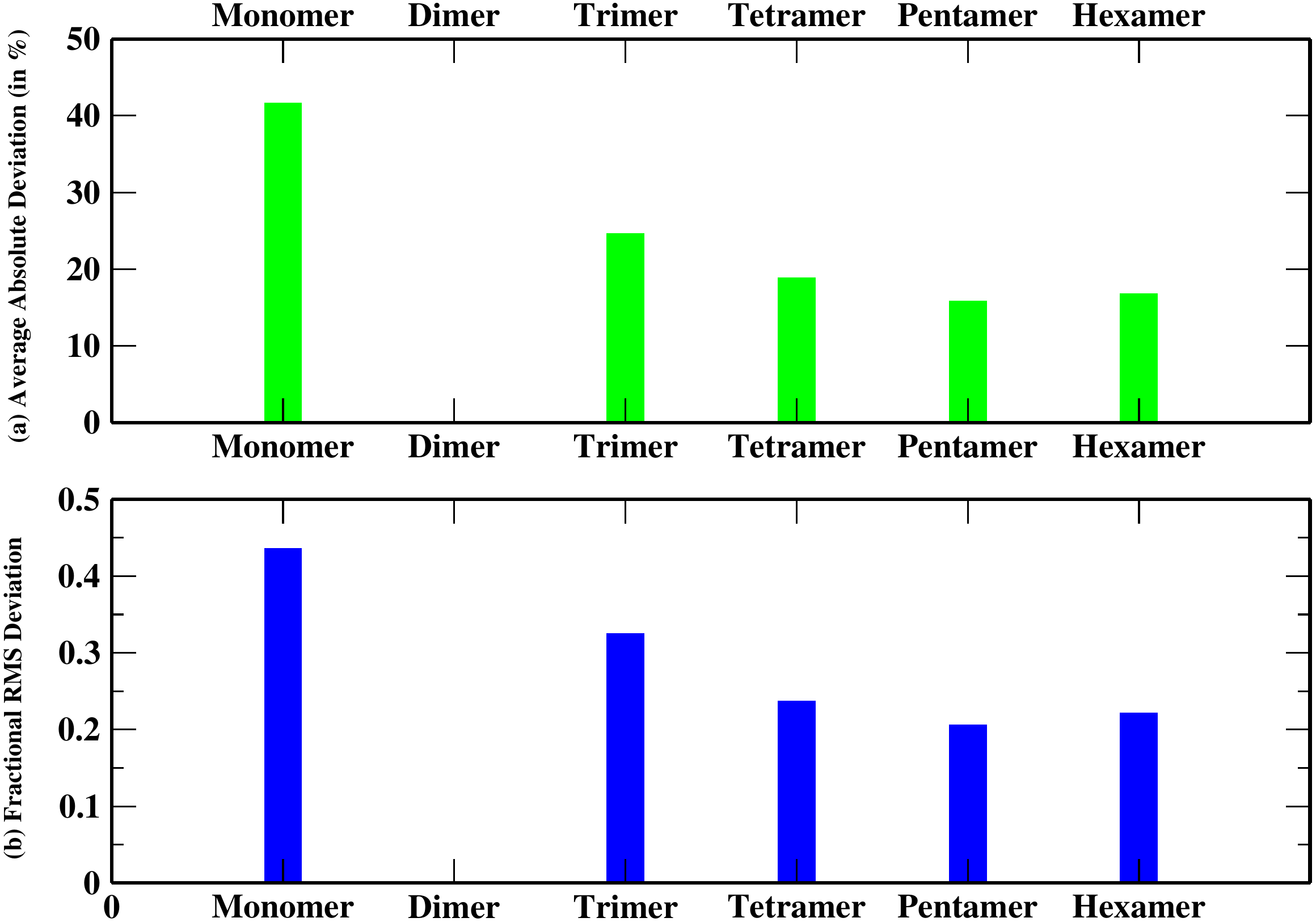}
\caption{(a) Average absolute percentage deviation and (b) fractional RMS deviation of our calculated values from experiment.}
\end{figure}

BE calculations with water pentamer and hexamer configurations is very time consuming and thus it is
difficult to apply for a large set of species. In Appendix (Table A1), we provide our calculated BE values
for 21 species with c-hexamer (chair) configuration. Since BE values taking water c-hexamer (chair) cluster configuration 
deviate by $\sim \pm 16.7 \%$, they can be scaled by $1 \pm 0.167$.
We also provide our calculated BE values for a large set of $100$ important
interstellar or circumstellar species, where, we consider the c-tetramer configuration (as given in Fig. 1)
of water cluster as a substrate (which deviates the experimental values by $\sim \pm 18.8 \%$) and
BE values of these species can also be scaled by $1 \pm 0.188$.
We think that our calculations for the tetramer is an interesting
alternative to full calculation or to the fitting model used by \cite{wake17}. Some strange values of BE of some species
are due to induce deviation for the global minimum.
The mobility of atoms is very important for grain chemistry in dense molecular clouds which depend mostly
on their adsorption energy with grain surface. BEs of some relevant atoms (such as H, He, C, N, O, Na, Mg, Si, P, and S) 
are also included in Table A1. With the tetramer configuration, our calculations likely deviate the experimental BE values 
due to the long-range interaction (interaction with water molecule not close to the species as visible from
Fig. A3(a-e) which seems crucial for the cases of low BEs. 
So, using only a scale factor without offset with the tetramer configuration may 
likely underestimate the real BE, particularly for low BE cases. \\

Previously used and recently proposed KIDA \url{(http://kida.obs.u-bordeaux1.fr)} BE values by \cite{wake17} and 
BE values provided by other sources are given in columns 6, 7 and 8 respectively of Table A1.
Among the sixteen stable species given in Table 1 and 2, there are both polar 
(OCS, CH$_3$OH, NH$_3$, H$_2$S, CH$_3$CCH, CH$_3$CN, H$_2$O$_2$, HCl, HNCO, NO) and non-polar molecules 
(CO, CO$_2$, N$_2$, O$_2$, CH$_4$, C$_2$H$_2$).We have noticed that for the non-polar species if we choose the position
at the center of water cluster, our calculation sometimes overestimates the attractive interaction. For species (like HCl, CH$_3$OH, etc.) leading to hydrogen bond with H$_2$O, the species is rather localized
on one water molecule with H-bonding and in most of the cases, the interaction is well reproduced by our calculations.
We found the case of oxygen atom as an intermediate one. We have obtained the BE of oxygen atom 
to be $1002$ K and $660$ K (given in Table A1) respectively with the tetramer and c-hexamer (chair) configuration whereas
the existing value was $1660 \pm 60$ K in KIDA desorption energy on ASW surface. This is because of the complex containing
water c-hexamer (chair) with oxygen atom is notably modified and so the interaction between water is less good and it seems
that this is not compensated by the interaction with the oxygen atom. But in the case of the complex containing water tetramer
with oxygen atom gives more accurate value likely because in that case water interaction is not so perturbed. 
The hydrogen atom case ($125$ K with tetramer and $181$ K with hexamer from Table A1) is a very strange one as it leads
to a much smaller interaction in comparison to the value ($= 650$ K) mentioned in \cite{alha07}.
Thus, though we are having very good approximation (on an average) by using the c-pentamer and
c-hexamer (chair) configuration, 
we should keep in mind that using only one adsorption site geometry may induce notable error on BE.
For a given method and for a given species, it is recommended to consider various adsorption sites and
various stable geometries and take an average value to come up with a more relevant approximation. 
In our case, when various binding sites where found, we try to choose the closest one with previous calculations.
In Figs. A1, A2, and A3(a-e) of the appendix, we provide our optimized geometries for the species considered here
with c-hexamer (chair), c-pentamer, and c-tetramer configurations
respectively. Used ground states of each species are also provided in column 3 of Table A1. \\

\section{Conclusions}
In this work, we compute BE of various interstellar species on water surface. We performed a systematic
quantum chemical calculations for various adsorbed species by considering water monomer, 
c-trimer, c-tetramer, c-pentamer, and c-hexamer (chair) configuration respectively as a substrate
to investigate the physisorption BE of species. The advantage of using two or more water molecules
is that one can take into account the H bonding donor and acceptor behaviour of some molecules.
We noticed that our calculated values of BE tend to experimental values with increasing
cluster size of water molecules which is very clear from calculated absolute average percentage deviation
and root mean square deviation values given in Table 1.  More specifically, when we considered the c-pentamer and
c-hexamer (chair) configurations, on an average, our computed BE values deviate with experimental values within of
$\sim \pm 15.8 \%$ and $\sim \pm 16.7 \%$ respectively. The results obtained using a low level of theory (small basis set, 
no BSSE correction, no ZPE) is in better agreement with experiment than a higher level likely due to the fact that the various
approximations at low level compensate each other. This is probably a coincidence, but is useful as it makes it easier to process a large
number of systems. For a wide set of interstellar or circumstellar species, we provide the BE values with the c-tetramer configuration 
which are deviated by on an average of $\sim \pm 18.8 \%$. 
These comparisons show that we can safely use our procedure to compute BE of any such molecules with reasonable accuracy.

\section{Acknowledgement}
MS gratefully acknowledges DST, the Government of India for providing financial assistance through DST-INSPIRE Fellowship
[IF160109] scheme. AD and PG acknowledge ISRO respond project  (Grant No. ISRO/RES/2/402/16-17) for partial financial support. \\

\software{Gaussian 09 \citep{fris13}}

\clearpage

{\hskip 7.5cm \Large \bf Appendix}\\\\
\noindent {{\bf Table A1:} Calculated and Available list of} BEs of Various Species.
\vskip 0.1cm

\begin{table}
\scriptsize{
\centering
\hskip -2.9cm
\begin{tabular}{||c|c|c|c|c|c|c|c||}
\hline
\hline
{\bf Serial} & {\bf Species} & {\bf Ground } & {\bf BE in Kelvin } & {\bf BE in Kelvin} & \multicolumn{2}{|c|}{\bf \underline{BE from KIDA database}} & {\bf BE in Kelvin} \\
{\bf Number} &&{\bf state used} & {\bf on water tetramer}&{\bf on water hexamer}  & {\bf Old values} & {\bf New values} & {\bf  from other} \\
& &  & & & {\bf (Kelvin)} & {\bf (Kelvin)} & {\bf literature sources} \\
\hline
\hline
1. & H & Doublet & 125 & 181 & 450$^g$ & 650 $\pm$ 195 & 650 $\pm$ 100$^c$ \\
2. & H$_2$ & Singlet & 528 & 545 & 450 & 440 $\pm$ 132 & 500 $\pm$ 100$^c$ \\
3. & He & Singlet & 113 &  & 100 & & 100 $\pm$ 50$^c$  \\
4. & C & Triplet & 660 &  & 800 & 10000 $\pm$ 3000 & 715 $\pm$ 360$^c$, 14100 $\pm$ 420$^k$ \\
5. & N & Quartet & 269 & 619 & 800 & 720 $\pm$ 216 & 715 $\pm$ 358$^c$, 400 $\pm$ 30$^k$ \\
6. & O & Triplet & 1002 & 660 & 1660 $\pm$ 60$^a$ & 1600 $\pm$ 480 & 1660 $\pm$ 60$^c$, 1504 $\pm$ 12$^j$, 1440 $\pm$ 160$^k$ \\
7. & Na & Doublet & 2214  && 11800 & & 10600 $\pm$ 500$^c$ \\
8. & Mg & Singlet & 654 && 5300 & & 4750 $\pm$ 500$^c$ \\
9. & Si & Triplet & 6956 && 2700 & 11600 $\pm$ 3480 & 2400 $\pm$ 500$^c$ \\
10. & P & Quartet & 616 && 1100 & & 750 $\pm$ 375 \\
11. & S & Triplet & 1428 && 1100 & 2600 $\pm$ 780 & 985 $\pm$ 495$^c$  \\
12. & NH & Triplet & 1947 && 2378 & 2600 $\pm$ 780 & 542 $\pm$ 270$^c$ \\
13. & OH & Doublet & 3183 && 2850$^g$ & 4600 $\pm$ 1380 & 3210 $\pm$ 1550$^c$ \\
14. & PH & Triplet & 944 && && 800 $\pm$ 400 \\
15. & C$_2$ & Triplet & 9248 && 1600 & & 1085 $\pm$ 500$^c$ \\
16. & HF & Singlet & 5540 &&&& 500 $\pm$ 250 \\
17. & HCl & Singlet & 3924 & 4104 & 5174 $\pm$ 1$^b$ & 5172 $\pm$ 1551.6 & 900 $\pm$ 450 \\
18. & CN & Doublet & 1736 & & 1600 & & 1355 $\pm$ 500$^c$ \\
19. & N$_2$ & Singlet & 900 & 1161 & 1000$^g$ & 1100 $\pm$ 330 & 990 $\pm$ 100$^c$, 1000$^f$ \\
20. & CO & Singlet & 1263 & 1292 & 1150$^g$ & 1300 $\pm$ 390 & 1100 $\pm$ 250$^c$, 1300$^e$ \\
21. & SiH & Doublet & 8988 & & 3150 & 13000 $\pm$ 3900 & 2620 $\pm$ 500$^c$  \\
22. & NO & Doublet & 1265 & 1988 & 1600 & 1600 $\pm$ 480 & 1085 $\pm$ 500$^c$ \\
23. & O$_2$ & Triplet & 940 & 1352 & 1000 & 1200 $\pm$ 360 & 898 $\pm$ 30$^c$, 1200$^e$, 1000$^f$ \\
24. & HS & Doublet & 2221 & & 1450 & 2700 $\pm$ 810 & 1350 $\pm$ 500$^c$ \\
25. & SiC & Triplet & 5850 & & 3500 & & 3150 $\pm$ 500$^c$ \\
26. & CP & Doublet & 1699  & & 1900 & & 1050 $\pm$ 500 \\
27. & CS & Singlet & 2217  && 1900 & 3200 $\pm$ 960 & 1800 $\pm$ 500$^c$ \\
28. & NS & Doublet & 2774 && 1900 & & 1800 $\pm$ 500$^c$ \\
29. & SO & Triplet & 2128  && 2600 & 2800 $\pm$ 840 & 1800 $\pm$ 500$^c$ \\
30. & S$_2$ & Triplet & 1644 && 2200 & & 2000 $\pm$ 500$^c$ \\
31. & CH$_2$ & Triplet & 1473 && 1050 & 1400 $\pm$ 420 & 860 $\pm$ 430$^c$ \\
32. & NH$_2$ & Doublet & 3240  && 3956 & 3200 $\pm$ 960 & 770 $\pm$ 385$^c$ \\
33. & H$_2$O & Singlet & 2670  & 4166 & 5700$^g$ & 5600 $\pm$ 1680 & 4800 $\pm$ 100$^c$ \\
34. & PH$_2$ & Doublet & 1226  && 2000 && 850 $\pm$ 425 \\
35. & C$_2$H & Doublet & 2791 && 2137 & 3000 $\pm$ 900 & 1330 $\pm$ 500$^c$ \\
36. & N$_2$H & Doublet & 3697 && 1450 & & \\
37. & O$_2$H & Doublet & 5778 && 3650 & 5000 $\pm$ 1500 & 1510 $\pm$ 500$^c$ \\
38. & HS$_2$ & Doublet & 4014 && 2650 & & 2300 $\pm$ 500$^c$ \\
39. & HCN & Singlet & 2352 && 2050 & 3700 $\pm$ 1110 & 1580 $\pm$ 500$^c$ \\
40. & HNC & Singlet & 5225 && 2050 & 3800 $\pm$ 1140 & 1510 $\pm$ 500$^c$ \\
41. & HCO & Doublet & 1857 && 1600$^g$ & 2400 $\pm$ 720 & 1355 $\pm$ 500$^c$ \\
42. & HOC & Doublet & 5692 && & & \\
43. & HCS & Doublet & 2713 && 2350 & 2900 $\pm$ 870 & 2000 $\pm$ 500$^c$ \\
44. & HNO & Singlet & 2988 && 2050 & 3000 $\pm$ 900 & 1510 $\pm$ 500$^c$ \\
45. & H$_2$S & Singlet & 2556 & 3232 & 2743$^f$ & 2700 $\pm$ 810 & 2290 $\pm$ 90$^c$ \\
46. & C$_3$ & Singlet & 2863 && 2400 & 2500 $\pm$ 750 & 2010 $\pm$ 500$^c$ \\
47. & O$_3$ & Singlet & 2545 && 1800 & 2100 $\pm$ 630 & 2100 $\pm$ 100$^c$ \\
48. & C$_2$N  & Doublet & 1281 && 2400 & & 2010 $\pm$ 500$^c$ \\
49. & C$_2$S  & Triplet & 2477 && 2700 & & 2500 $\pm$ 500$^c$ \\
50. & OCN & Doublet & 3085 && 2400 & & 1805 $\pm$ 500$^c$ \\

\hline
\hline
\end{tabular}}
\end{table}

\begin{table}
\scriptsize{
\hskip -3cm
\begin{tabular}{||c|c|c|c|c|c|c|c||}
\hline
\hline
{\bf Serial} & {\bf Species} & {\bf Ground} & {\bf BE in Kelvin } & {\bf BE in Kelvin} & \multicolumn{2}{|c|}{\bf \underline{BE from KIDA database}} & {\bf BE in Kelvin} \\
{\bf Number} & & {\bf state used} & {\bf on water tetramer} & {\bf on water hexamer}  & {\bf Old values} & {\bf New values} & {\bf from other} \\
& & & & & {\bf (Kelvin)} & {\bf (Kelvin)} & {\bf literature sources}\\
\hline
\hline
51. & CO$_2$ & Singlet & 2293 & 2352 & 2575 & 2600 $\pm$ 780 & 2267 $\pm$ 70$^c$, 2300$^e$ \\
52. & OCS & Singlet & 1571 & 1808 & 2888 & 2400 $\pm$ 720 & 2325 $\pm$ 95$^c$ \\
53. & SO$_2$ & Singlet & 3745 && 3405 & 3400 $\pm$ 1020 & 3010 $\pm$ 110$^c$ \\
54. & CH$_3$ & Doublet & 1322  && 1175 & 1600 $\pm$ 480 & 1040 $\pm$ 500$^c$ \\
55. & NH$_3$ & Singlet & 3825  & 5163 & 5534 & 5500 $\pm$ 1650 & 2715 $\pm$ 105$^c$, 5530$^f$ \\
56. & SiH$_3$ & Doublet & 1269 && 4050 & & 3440 $\pm$ 500$^c$ \\
57. & C$_2$H$_2$ & Singlet & 2593 & 2640 & 2587$^f$ & 2587 $\pm$ 776.1 & 2090 $\pm$ 85$^c$ \\
58. & N$_2$H$_2$ & Singlet & 3183 &&& & \\
59. & H$_2$O$_2$ & Singlet & 3928 & 4248 & 5700 & 6000 $\pm$ 1800 & 6000 $\pm$ 100$^c$, 5410$^l$ \\
60. & H$_2$S$_2$ & Singlet & 4368 && 3100 & & 2600 $\pm$ 500$^c$ \\
61. & H$_2$CN & Doublet & 2984 && 2400 & & 2400 $\pm$ 500$^c$ \\
62. & CHNH & Doublet & 3742 &&& & \\
63. & H$_2$CO & Singlet & 3242 && 2050$^g$ & 4500 $\pm$ 1350 & 3260 $\pm$ 60$^c$  \\
64. & CHOH & Triplet & 4800 &&& & \\
65. & HC$_2$N & Triplet & 3289 &&& & 2270 $\pm$ 500$^c$ \\
66. & HC$_2$O & Doublet & 2914 && 2400 & & 2010 $\pm$ 500$^c$ \\
67. & HNCO & Singlet & 3922 & 5554 & 2850 & 4400 $\pm$ 1320 & 2270 $\pm$ 500$^c$, 3900$^h$ \\
68. & H$_2$CS & Singlet & 3110 && 2700 & 4400 $\pm$ 1320 & 2025 $\pm$ 500$^c$ \\
69. & C$_3$O & Singlet & 3542 && 2750 & & 2520 $\pm$ 500$^c$ \\
70. & CH$_4$ & Singlet & 1327 & 2321 & 1300$^g$ & 960 $\pm$ 288 & 1250 $\pm$ 120$^c$ \\
71. & SiH$_4$ & Singlet & 1527 && 4500 & & 3690 $\pm$ 500$^c$ \\
72. & C$_2$H$_3$ & Doublet & 2600 && 3037 & 2800 $\pm$ 840 & 1760 $\pm$ 500$^c$  \\
73. & CHNH$_2$ & Triplet & 1681 &&& & \\
74. & CH$_2$NH & triplet & 3354  && 5534 & & 1560 $\pm$ 500$^c$ \\
75. & CH$_3$N & Triplet & 2194 &&& & \\
76. & c-C$_3$H$_2$ & Singlet & 3892 && 3387 & 5900 $\pm$ 1770 & 2110 $\pm$ 500$^c$ \\
77. & H$_2$CCN & Doublet & 3730  && 4230 &  & 2470 $\pm$ 500$^c$ \\
78. & H$_2$CCO & Singlet & 2847 && 2200 & 2800 $\pm$ 840 & 2520 $\pm$ 500$^c$ \\
79. & HCOOH & Singlet & 3483 && 5570$^g$ & & 4532 $\pm$ 150$^c$ \\
80. & CH$_2$OH & Doublet & 4772 && 5084 & 4400 $\pm$ 1320 & 2170 $\pm$ 500$^c$ \\
81. & NH$_2$OH & Singlet & 4799 && & & 2770 $\pm$ 500$^c$ \\
82. & C$_4$H & Doublet & 2946 && 3737 & & 2670 $\pm$ 500$^c$ \\
83. & HC$_3$N & Singlet & 2925  && 4580 & & 2685 $\pm$ 500$^c$ \\
84. & HC$_3$O & Doublet & 2619 &&& & \\
85. & C$_5$ & Singlet & 2403 && 4000 & & 3220 $\pm$ 500$^c$ \\
86. & C$_2$H$_4$ & Singlet & 2052 && 3487 & 2500 $\pm$ 750 & 2010 $\pm$ 500$^c$ \\
87. & CH$_2$NH$_2$ & Doublet & 3831 && 5534$^d$ & & \\
88. & CH$_3$NH & Doublet & 3414 &&& & 1760 $\pm$ 500$^c$ \\
89. & CH$_3$OH & Singlet & 4368 & 4511 & 5534 & 5000 $\pm$ 1500 & 3820 $\pm$ 135$^c$, 5530$^f$ \\
90. & CH$_2$CCH & Doublet & 2726 & & 3837 & 3300 $\pm$ 990 & 3840 $\pm$ 500$^c$ \\
91. & CH$_3$CN & Singlet & 2838 & 3786 & 4680$^f$ & 4680 $\pm$ 1404 & 3790 $\pm$ 130$^c$ \\
92. & H$_2$C$_3$N & Doublet & 2637 &&& & \\
93. & H$_2$C$_3$O & Singlet & 3006 &&& & \\
94. & C$_6$ & Quintet & 3226 && 4800 & & 3620 $\pm$ 500$^c$ \\
95. & CH$_3$NH$_2$ & Singlet & 4434 && 6584 & & 5130 $\pm$ 500$^c$, 4269$^m$ \\
96. & C$_2$H$_5$ & Doublet & 1752 && 3937 & 3100 $\pm$ 930 & 2110 $\pm$ 500$^c$ \\
97. & CH$_3$CCH & Singlet & 2342 & 3153 & 4287 & 3800 $\pm$ 1140 & 4290 $\pm$ 500$^c$, 2500 $\pm$ 40$^i$ \\
98. & CH$_2$CCH$_2$ & Triplet & 2705 && & 3000 $\pm$ 900 & 4290 $\pm$ 500$^c$ \\
99. & CH$_3$CHO & Singlet & 3849 && 2450 & 5400 $\pm$ 1620 & 2870 $\pm$ 500$^c$ \\
100. & C$_7$ & Singlet & 4178 && 5600 & & 4430 $\pm$ 500$^c$ \\
\hline
\hline
\end{tabular}} \\
\\
{\bf Notes.} Average deviation from experiment in case of tetramer (column 4) and hexamer (column 5) are 
$\sim \pm 18.8 \%$ and $\sim \pm 16.7 \%$ respectively. \\
$^a$ \cite{he15},
$^b$ \cite{olan11},
$^c$ \cite{pent17},
$^d$ \cite{ruau15},
$^e$ \cite{mini16},
$^f$ \cite{coll04},
$^g$ \cite{garr06},
$^h$ \cite{nobl15},
$^i$ \cite{kimb14},
$^j$ \cite{ward12},
$^k$ \cite{shim18},
$^l$ \cite{lamb17}.
$^m$ \cite{chaa18}
\end{table}

\clearpage

\hskip 1cm {\bf Figure A1.} Optimized Geometries with the c-hexamer (chair) Configuration.
\begin{figure}
\centering
\includegraphics[height=13cm, width=18cm]{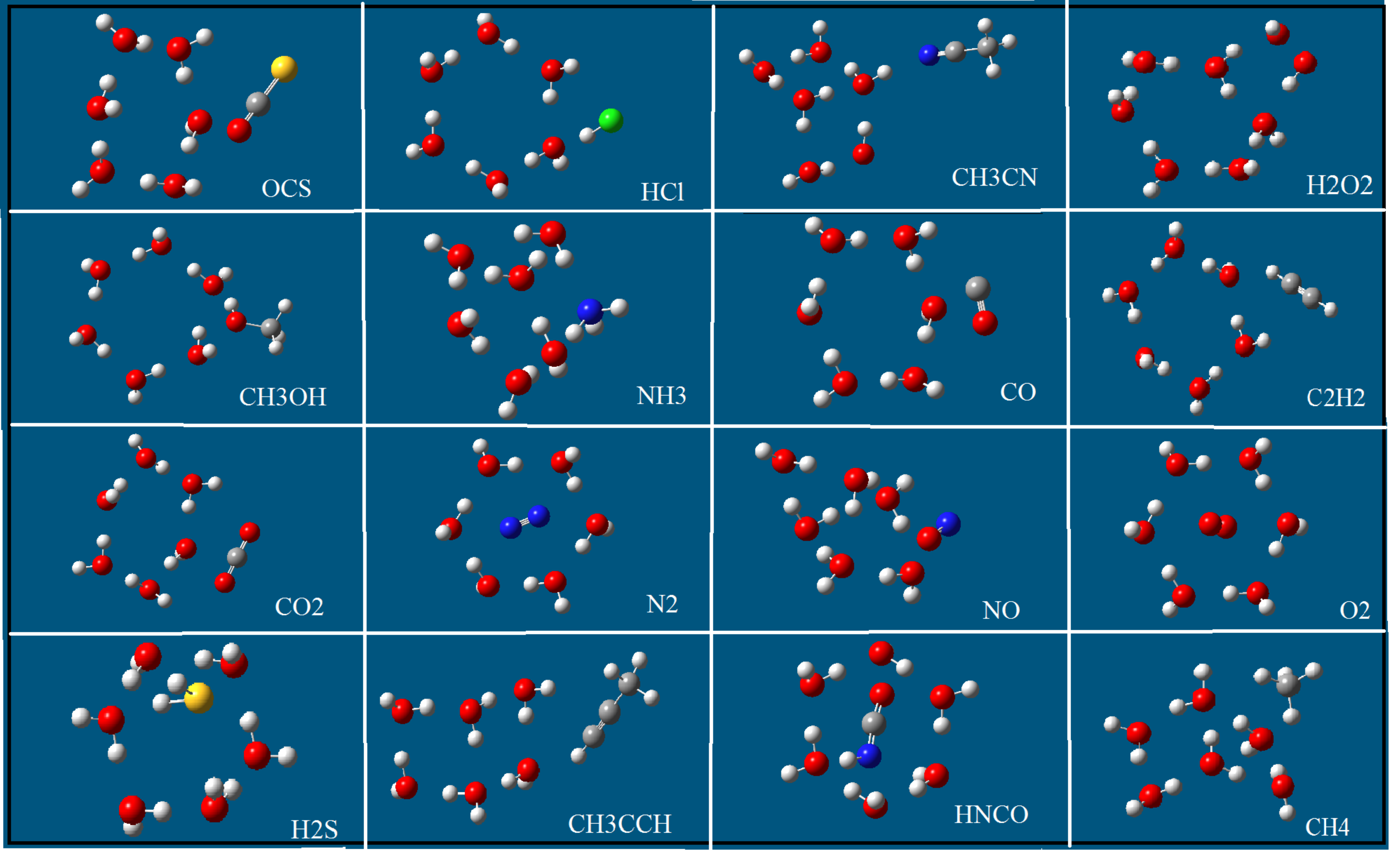}
\end{figure}
\clearpage

\hskip 1cm {\bf Figure A2.} Optimized Geometries with the c-pentamer Configuration.
\begin{figure}
\centering
\includegraphics[height=13cm, width=18cm]{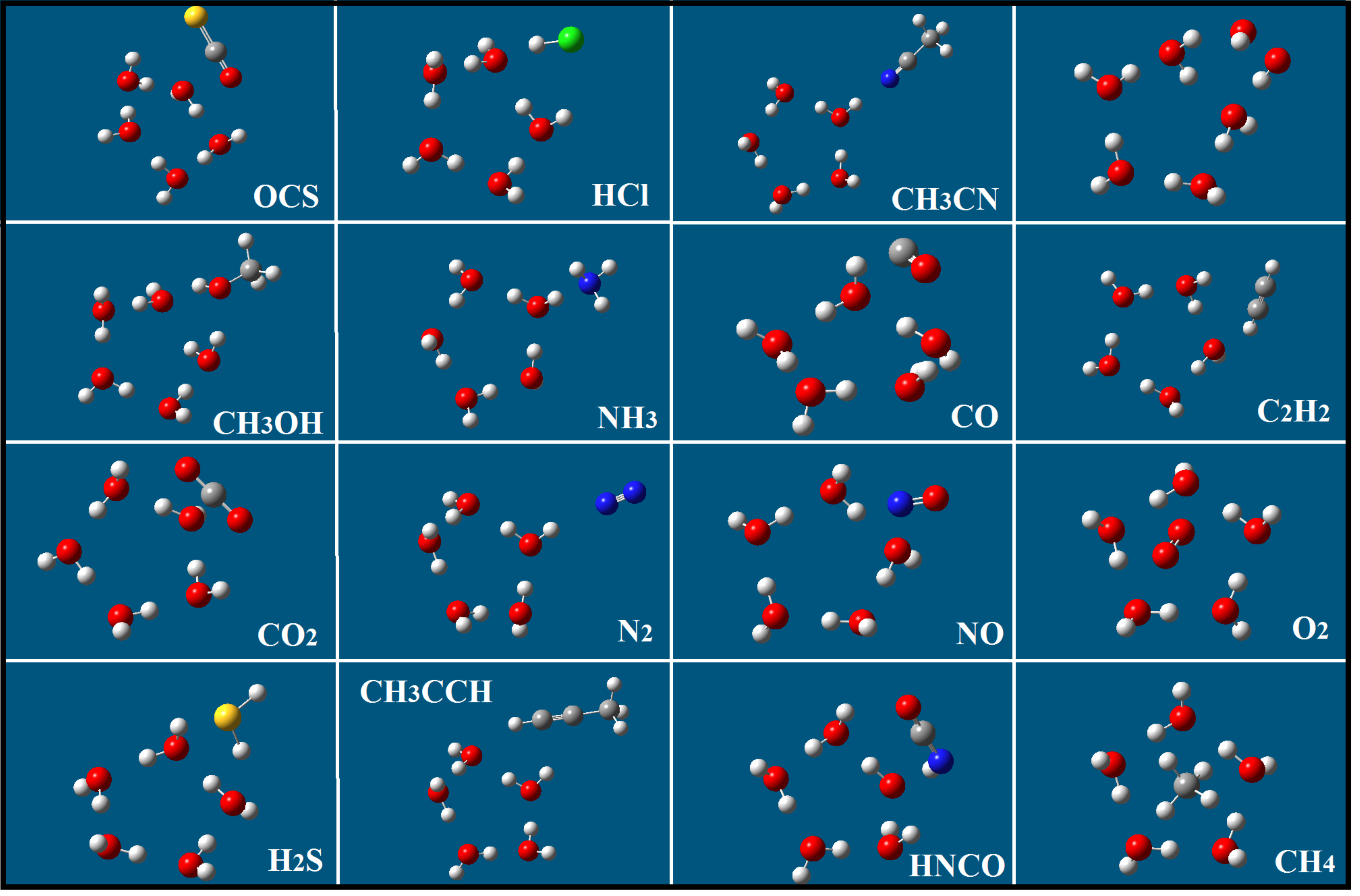}
\end{figure}
\clearpage

\hskip 1cm {\bf Figure A3(a).} Optimized Geometries with the c-tetramer Configuration.
\begin{figure}
\centering
\includegraphics[height=20cm, width=18cm]{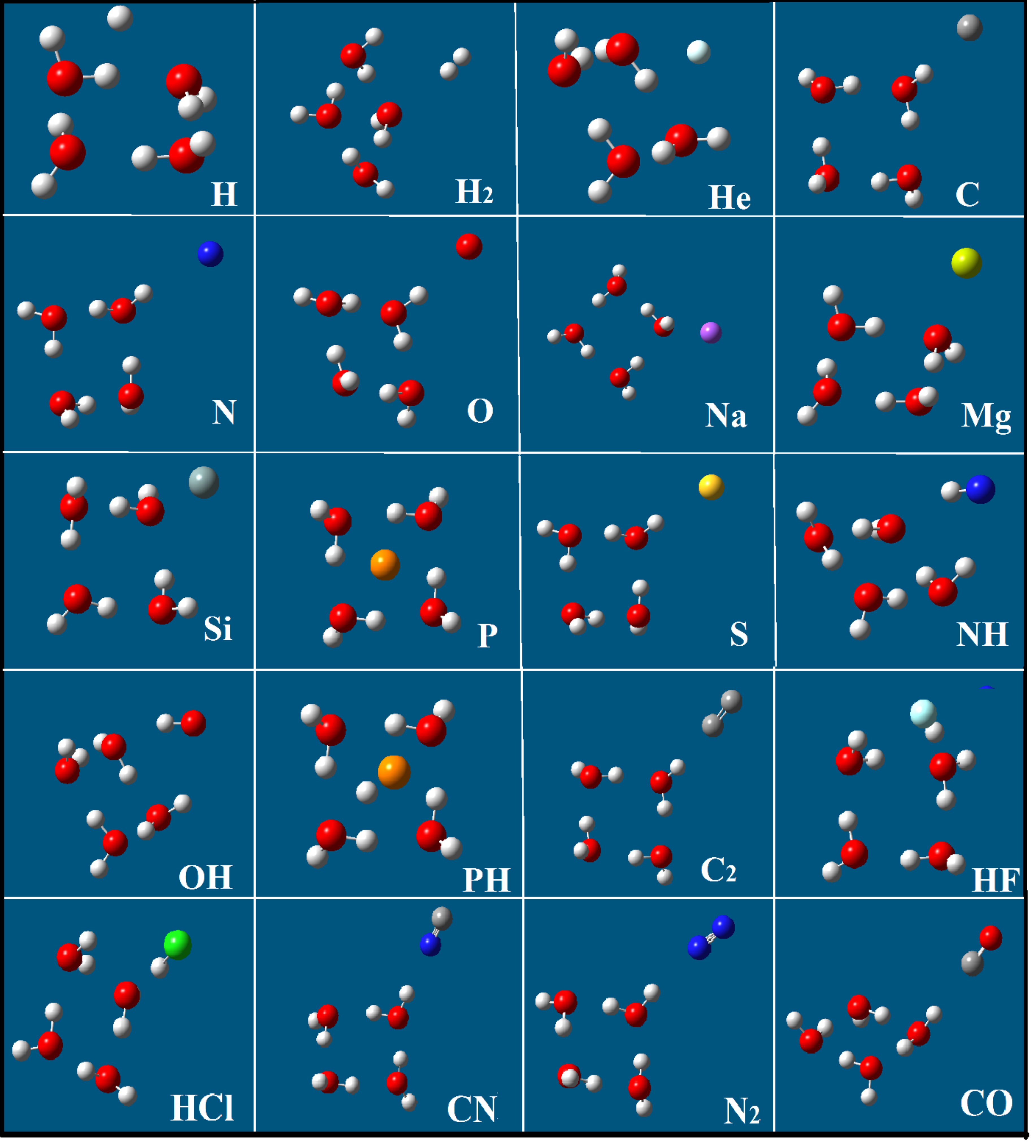}
\end{figure}
\clearpage

\hskip 1cm {\bf Figure A3(b).} Optimized Geometries with the c-tetramer Configuration.
\begin{figure}
\centering
\includegraphics[height=20cm, width=18cm]{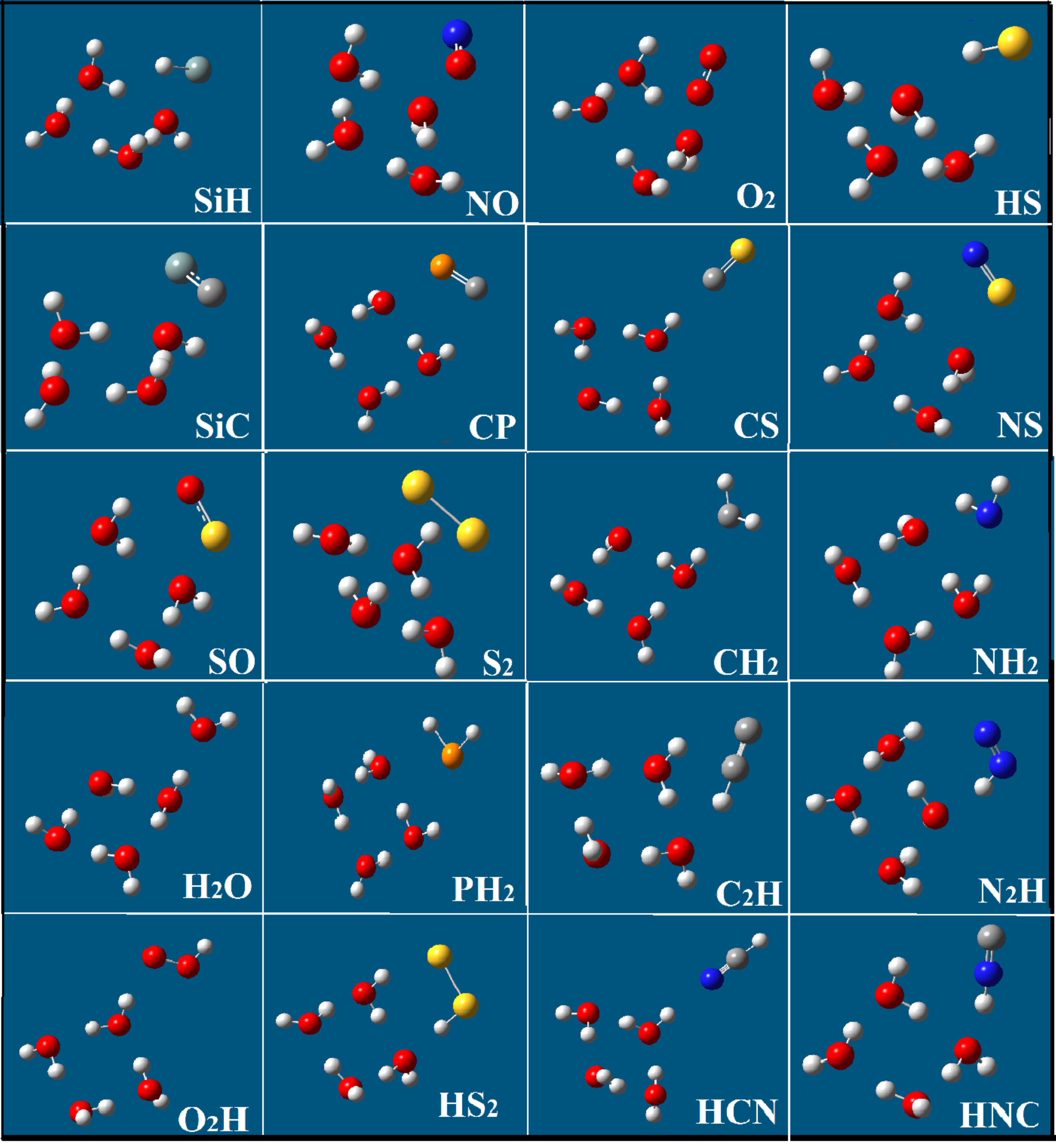}
\end{figure}
\clearpage

\hskip 1cm {\bf Figure A3(c).} Optimized Geometries with the c-tetramer Configuration.
\begin{figure}
\centering
\includegraphics[height=20cm, width=18cm]{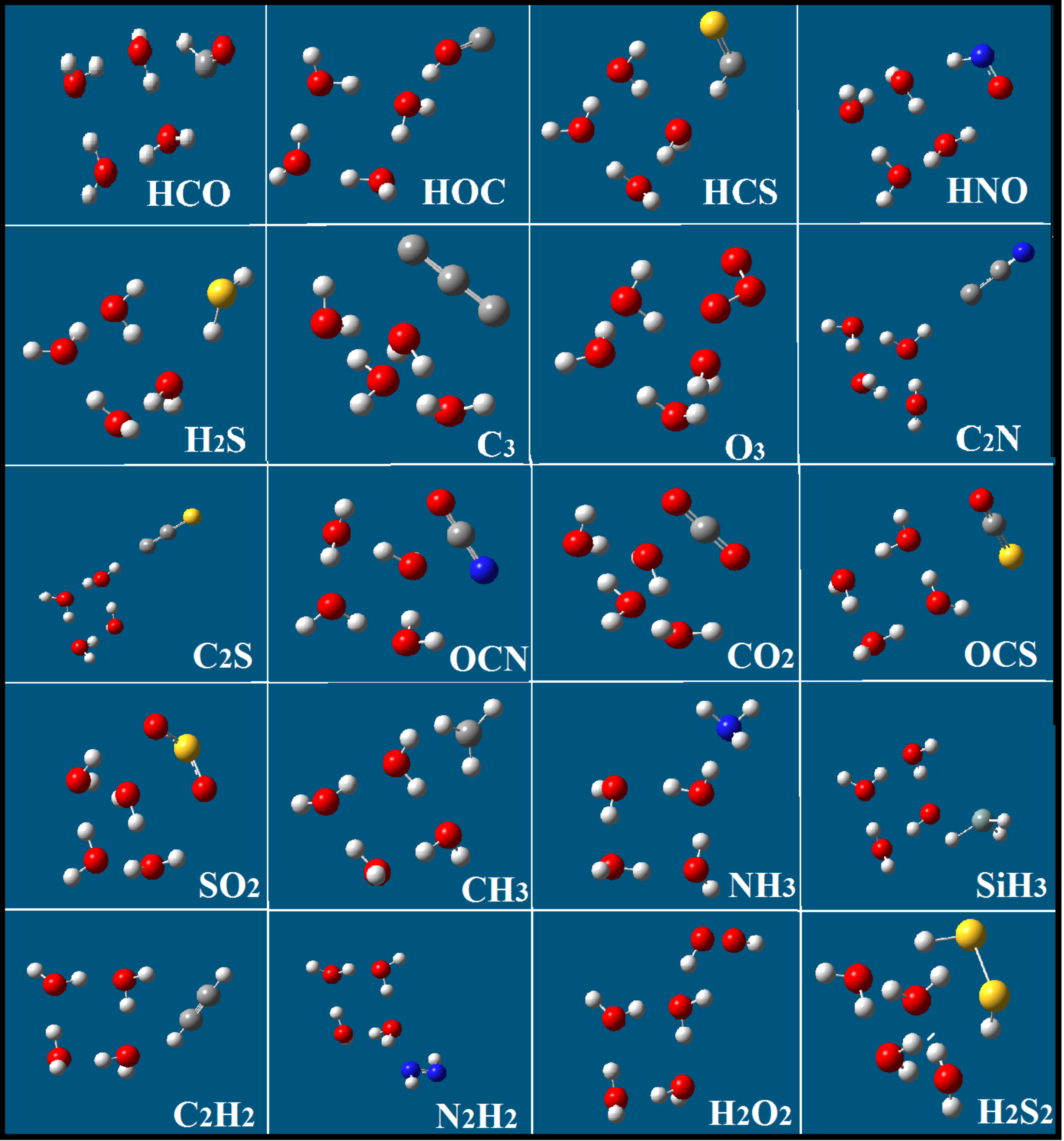}
\end{figure}
\clearpage

\hskip 1cm {\bf Figure A3(d).} Optimized Geometries with the c-tetramer Configuration.
\begin{figure}
\centering
\includegraphics[height=20cm, width=18cm]{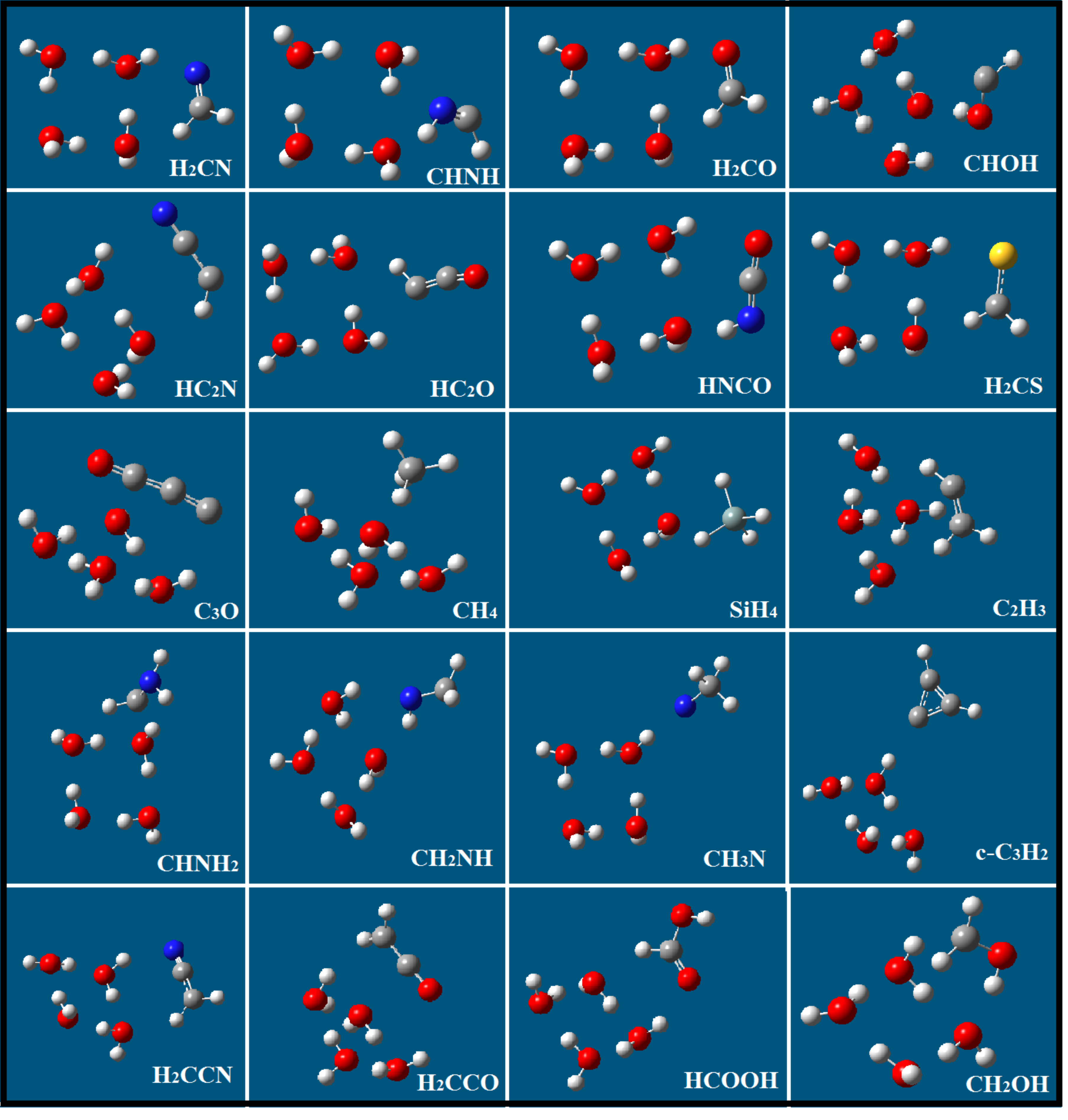}
\end{figure}
\clearpage

\hskip 1cm {\bf Figure A3(e).} Optimized Geometries with the c-tetramer Configuration.
\begin{figure}
\centering
\includegraphics[height=20cm, width=18cm]{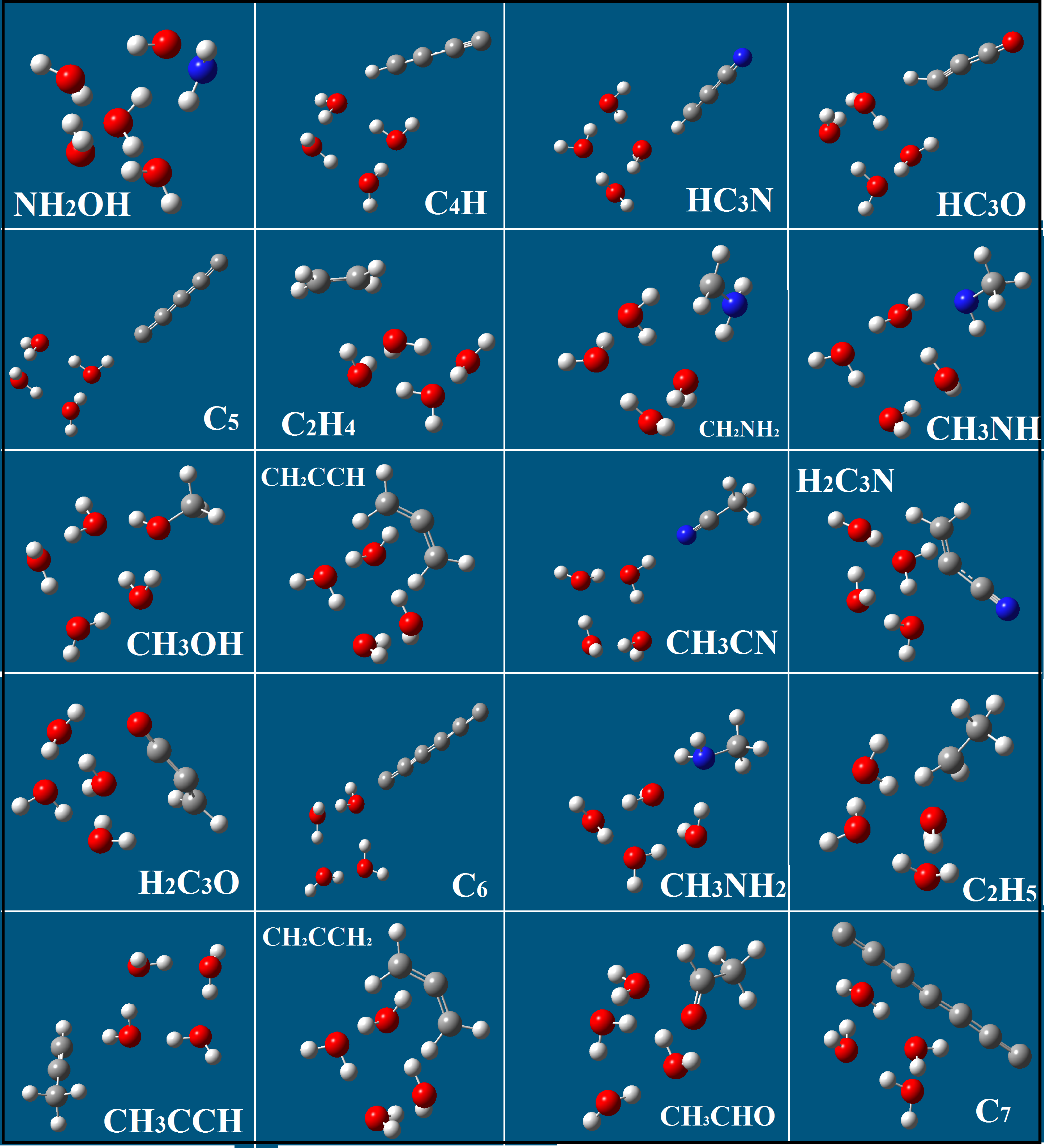}
\end{figure}
\clearpage


\clearpage
\begin{thebibliography}{}
\bibitem[\protect\citeauthoryear{Al-Halabi \& van Dishoeck}{2007}]{alha07}
Al-Halabi, A., \& van Dishoeck, E.F. 2007, MNRAS, 382, 1648
\bibitem[\protect\citeauthoryear{Chakarov \& Kasemo}{1998}]{chak98}
Chakarov, D., \& Kasemo, B. 1998, PhRvL, 81, 5181
\bibitem[\protect\citeauthoryear{Charnley}{1997}]{char97}
Charnley, S.B. 1997, MNRAS, 291, 455
\bibitem[\protect\citeauthoryear{Collings et al.}{2004}]{coll04}
Collings, M.P., Anderson, M.A., Chen, R., et al. 2004, MNRAS, 354, 1133
\bibitem[\protect\citeauthoryear{Chaabouni et al}{2018}]{chaa18}
Chaabouni, H., Diana, S., Nguyen, T., Dulieu, F. 2018, A\&A, doi: https://doi.org/10.1051/0004-6361/201731006
\bibitem[\protect\citeauthoryear{Das, Acharyya \& Chakrabarti}{2010}]{das10}
Das, A., Acharyya, K. \& Chakrabarti, S.K. 2010, MNRAS, 409, 789
\bibitem[\protect\citeauthoryear{Das \& Chakrabarti}{2011}]{das11}
Das, A. \& Chakrabarti, S.K. 2011, MNRAS, 418, 545
\bibitem[\protect\citeauthoryear{Das et al.}{2016}]{das16}
Das, A., Sahu, D., Majumdar, L., \& Chakrabarti, S.K. 2016, MNRAS, 455, 540
\bibitem[\protect\citeauthoryear{Dulieu et al.}{2013}]{duli13}
Dulieu, F., Congiu, E., Noble, J., et al. 2013, NatSR, 3, 1338
\bibitem[\protect\citeauthoryear{Dunning Jr}{1989}]{dunn89}
Dunning Jr, T.H. 1989, JChPh, 90(2), 1007
\bibitem[\protect\citeauthoryear{Frisch et al.}{2013}]{fris13}
Frisch, M. J., Trucks, G. W., Schlegel, H. B., et al. 2013, Gaussian 09, Revision D.01, Gaussian, Inc., Wallingford CT.
\bibitem[\protect\citeauthoryear{Garrod \& Herbst}{2006}]{garr06}
Garrod, R.T., \& Herbst, E. 2006, A\&A, 457, 927
\bibitem[\protect\citeauthoryear{Garrod et al}{2007}]{garr07}
Garrod, R.T., Wakelam, V., \& Herbst, E. 2007, A\&A, 467, 1103
\bibitem[\protect\citeauthoryear{Gorai et al.}{2017a}]{gora17a} 
Gorai, P., Das, A., Das, A., Sivaraman, B., et al. 2017, ApJ, 836,70
\bibitem[\protect\citeauthoryear{Gorai et al.}{2017b}]{gora17b} 
Gorai, P., Das, A., Majumdar. L., et al. 2017, MolAp, 6, 36-46
\bibitem[\protect\citeauthoryear{Hama \& Watanabe}{2013}]{hama13}
Hama, T., \& Watanabe, N. 2013, ChRv, 113, 8783
\bibitem[\protect\citeauthoryear{Hasegawa \& Herbst}{1993}]{hase93}
Hasegawa, T.I., \& Herbst, E. 1993, MNRAS, 263, 589
\bibitem[\protect\citeauthoryear{Herbst \& Van Dishoeck}{2009}]{herb09}
Herbst, E.; van Dishoeck, E. F. Annu. Rev. Astron. Astrophys. 2009, 47, 427
\bibitem[\protect\citeauthoryear{He et al.}{2015}]{he15}
He, J., Shi, J., Hopkins, T., Vidali, G., \& Kaufman, M.J. 2015, ApJ, 801(2), 120
\bibitem[\protect\citeauthoryear{Hornekær et al.}{2005}]{horn05}
Hornekær, L., Baurichter, A., Petrunin, V.V., et al. 2005, JChPh, 122(12), 124701
\bibitem[\protect\citeauthoryear{Karssemeijer \& Cuppen}{2014}]{kars14}
Karssemeijer, L.J., \& Cuppen, H.M. 2014, A\&A, 569, A107
\bibitem[\protect\citeauthoryear{Keane et al.}{2001}]{kean01}
Keane J.V., Tielens A.G.G.M., Boogert A.C.A., Schutte W.A., \& Whittet D.C.B. 2001, A\&A, 376, 254
\bibitem[\protect\citeauthoryear{Kimber et al.}{2014}]{kimb14}
Kimber, H.J., Ennis, C.P., \& Price, S.D. 2014, FaDi, 168, 167
\bibitem[\protect\citeauthoryear{Lamberts et al.}{2017}]{lamb17}
Lamberts, T., \& K$\ddot{a}$stner, J. 2017, ApJ, 846(1), 43
\bibitem[\protect\citeauthoryear{Maldoni et al.}{2003}]{mald03}
Maldoni, M.M., Egan, M.P., Smith, R.G., Robinson, G., \& Wright, C.M. 2003, MNRAS, 345, 912
\bibitem[\protect\citeauthoryear{Malfait et al.}{1998}]{malf98}
Malfait, K., Waelkens, C., Waters, L.B.F.M., et al. 1998, A\&A, 332, L25
\bibitem[\protect\citeauthoryear{Minissale \& Dulieu}{2014}]{mini14}
Minissale, M., \& Dulieu, F. 2014, JChPh, 141, 014304
\bibitem[\protect\citeauthoryear{Minissale et al.}{2016}]{mini16}
Minissale, M., Dulieu, F., Cazaux, S., \& Hocuk, S. 2016, A\&A, 585, A24
\bibitem[\protect\citeauthoryear{Noble et al.}{2012}]{nobl12}
Noble, J. A., Congiu, E., Dulieu, F., \& Fraser, H. J. 2012, MNRAS, 421(1), 768
\bibitem[\protect\citeauthoryear{Noble et al.}{2015}]{nobl15}
Noble, J.A., Theule, P., Congiu, E., et al. 2015, A\&A, 576, A91
\bibitem[\protect\citeauthoryear{Ohno et al.}{2005}]{ohno05}
Ohno, K., Okimura, M., Akaib, N., \& Katsumotoa, Y. 2005, PCCP, 7, 3005
\bibitem[\protect\citeauthoryear{Olanrewaju et al.}{2011}]{olan11}
Olanrewaju, B.O., Herring-Captain, J., Grieves, G.A., Aleksandrov, A., \& Orlando, T.M. 2011, JPCA, 115, 5936
\bibitem[\protect\citeauthoryear{Palumbo}{2006}]{palu06}
Palumbo, M.E. 2006, A\&A, 453, 903
\bibitem[\protect\citeauthoryear{Penteado et al.}{2017}]{pent17}
Penteado, E.M., Walsh, C., \& Cuppen, H.M. 2017, ApJ, 844(1), 71
\bibitem[\protect\citeauthoryear{Raut et al.}{2007}]{raut07}
Raut, U., Famà, M., Teolis, B.D., et al. 2007, JChPh, 127, 204713
\bibitem[\protect\citeauthoryear{Ruaud et al.}{2015}]{ruau15}
Ruaud, M., Loison, J.C., Hickson, K.M., et al. 2015, MNRAS, 447(4), 4004
\bibitem[\protect\citeauthoryear{Shimonishi et al.}{2018}]{shim18}
Shimonishi, T., Nakatani, N., Furuya, K., \& Hama, T. 2018, arXiv preprint, arXiv:1801.08716.
\bibitem[\protect\citeauthoryear{Sil et al.}{2018}]{sil18}
Sil, M., Gorai, P., Das, A., et al. 2018, ApJ, 853, 2,
\bibitem[\protect\citeauthoryear{Sil et al.}{2017}]{sil17}
Sil, M., Gorai, P., Das, A., Sahu, D., \& Chakrabarti, S.K. 2017, EPJD, 71, 45
\bibitem[\protect\citeauthoryear{Song \& K$\ddot{a}$stner}{2016}]{song16}
Song, L., \& K$\ddot{a}$stner, J. 2016, PCCP, 18(42), 29278
\bibitem[\protect\citeauthoryear{Tielens \& Hagen}{1982}]{tiel82}
Tielens, A.G.G.M., \& Hagen, W. 1982, A\&A, 114, 245
\bibitem[\protect\citeauthoryear{Wakelam et al.}{2017}]{wake17}
Wakelam, V., Loison, J.-C., Mereau, R., \& Ruaud, M. 2017, MolAs, 6, 22
\bibitem[\protect\citeauthoryear{Ward et al.}{2012}]{ward12}
Ward, M.D., Hogg, I.A., \& Price, S.D. 2012, MNRAS, 425, 1264
\bibitem[\protect\citeauthoryear{Williams \& Herbst}{2002}]{will02}
Williams, D. \& Herbst, E. 2002, SurSc, 500, 823
\bibitem[\protect\citeauthoryear{Whittet et al.}{1988}]{whit88}
Whittet, D.C.B., Bode, M.F., Longmore, A.J., et al. 1988, MNRAS, 233, 321
\end{thebibliography}
\end{document}